\documentclass[
  aps,
  prd,
  reprint,
  superscriptaddress,
  amsmath,
  amssymb,
  longbibliography,
  floatfix
]{revtex4-2}

\usepackage{graphicx}
\usepackage{booktabs}

\newcommand{\dd}{\mathrm{d}}
\newcommand{\ii}{\mathrm{i}}
\newcommand{\Order}{\mathcal{O}}
\newcommand{\Lam}{\Lambda}
\newcommand{\dlt}{\delta}
\newcommand{\vev}{v}
\newcommand{\Lzero}{L_0}
\newcommand{\Pbar}{\overline{P}}

\begin{document}

\title{Open-Channel Radiation Zeros and Nonlinear Damping of a Critical-Bubble Internal Mode}

\author{Tomohiro Inagaki}
\email[Contact author: ]{inagaki@hiroshima-u.ac.jp}
\affiliation{Information Media Center, Hiroshima University, Higashi-Hiroshima 739-8511, Japan}
\author{Yuko Murakami}
\affiliation{Information Media Center, Hiroshima University, Higashi-Hiroshima 739-8511, Japan}
\date{\today}

\begin{abstract}
We study radiation and nonlinear damping of a spherically symmetric localized mode of an $O(3)$ thermal critical bubble. The leading second-harmonic radiation amplitude is determined by the overlap between a nonlinear source and a continuum scattering wave. This overlap can vanish through destructive interference as the supercooling is varied, even when the second harmonic lies above the continuum threshold.
A signed-overlap scan resolves five zeros in a finite parameter interval, including three on
the thinner-wall side of the two representative roots studied in detail.
For these representative roots, half-line Jost--Green calculations,
threshold-preserving wall deformations, and constrained radial evolution
verify the cancellation and its robustness.  Near a selected root,
fourth-order perturbation theory predicts an $A^2$ displacement of the
finite-amplitude radiation minimum and an $A$-linear width of the
third-harmonic-dominated region; radial evolution quantitatively tests both
relations.  At the selected zero, third-harmonic loss predicts $A\sim\tau^{-1/4}$,
instead of the generic $A\sim\tau^{-1/2}$ law.  Measured harmonic powers and
a slow-envelope reconstruction support this hierarchy.  Independent
center-stable shooting recovers the unprojected third-harmonic coefficient,
which differs from the fixed-projector value.  The results connect a tunable
radiation form factor to nonlinear relaxation of an internal mode on an
unstable nucleation saddle.

\end{abstract}

\maketitle

\section{Introduction}
\label{sec:intro}

First-order phase transitions proceed by nucleation of critical bubbles that
separate a stable phase from a metastable background.  Semiclassically, the
nucleation saddle is characterized not only by its Euclidean action but also
by the spectrum of its quadratic fluctuation operator
\cite{Coleman1977,CallanColeman1977,Linde1983}.  For a thermal transition the
dominant saddle is often $O(3)$ symmetric.  Its Hessian contains the familiar
negative mode and translational zero modes, but it can also support positive
localized excitations of the wall.  Localized modes are known to influence
the subsequent nonlinear evolution of solitons and unstable saddles
\cite{MantonMerabet1997,MantonRomanczukiewicz2023,NavarroObregonQueiruga2024};
for a critical bubble, they connect the static spectroscopy of the
nucleation saddle to its Lorentzian dynamics.

The mode considered here is the spherical continuation of the shape
excitation of a planar quartic kink.  In the degenerate planar limit the
fluctuation operator reduces to the modified P\"oschl--Teller problem, with a
localized shape eigenvalue below the bulk continuum
\cite{PoschlTeller1933,Rajaraman1982}.  Curvature-induced excitation of the
same degree of freedom has also been studied for spherical $\phi^4$ walls
\cite{MalomedMaslov1991}.  In Ref.~\cite{InagakiMurakamiLocalized2026} we
followed the corresponding shape-connected branches of the $O(3)$ critical
bubble through finite supercooling and determined their normalized
wave functions and continuum thresholds.  Here we study how the spherically
symmetric localized mode loses energy to the surrounding false vacuum when
it is excited.

For a generic localized nonlinear oscillator, radiation begins at the first
nonlinear harmonic that enters the continuum.  The wobbling $\phi^4$ kink is
the standard example: second-harmonic emission produces algebraic damping
\cite{MantonMerabet1997,BarashenkovOxtoby2009}.  Internal-mode radiation has
also been quantified for global and Abelian--Higgs vortices
\cite{BlancoPilladoEtAl2021,AlonsoIzquierdoEtAl2024}.  The coupling to an
on-shell continuum state is encoded in a nonlinear Fermi--golden--rule (FGR)
coefficient.  Cubic coupling leads to the familiar inverse-fourth-root
amplitude law \cite{SofferWeinstein1999,AnSoffer2020}, while a nonzero
quadratic FGR coefficient yields inverse-square-root decay
\cite{LegerPusateri2026}.  Higher-order damping in a weak-resonance regime
can instead result from lower harmonics lying below the continuum threshold
\cite{LeiLiuYang2022}.  Here the second harmonic is already propagating:
the quantity that vanishes is its source overlap.  Vanishing leading FGR
constants also arise in linear metastability theory
\cite{CorneanJensenNenciu2015}; that setting provides a useful analogy but
does not establish the nonlinear decay law of the present saddle.

Destructive radiation minima have direct precedents in oscillon physics
\cite{FodorEtAl2009,SalmiHindmarsh2012,ZhangEtAl2020,CyncynatesGiurgicaTiron2021}.
In particular, harmonic-resolved calculations show that a suppressed leading
harmonic can leave a higher harmonic to control the loss
\cite{ZhangEtAl2020}.  The coherent-source and Green-function analysis of
Ref.~\cite{CyncynatesGiurgicaTiron2021} also makes the role of spatial phase
cancellation explicit.  We extend this mechanism to a normalized linear
Hessian eigen-mode of a static nucleation saddle.  Supercooling controls the
bubble profile, the internal eigenfunction, and the scattering phase, allowing
an open-channel radiation zero to be located spectrally and continued into a
finite-amplitude radiation minimum.

The unstable direction of the critical bubble adds a separate dynamical
question.  A fixed projection isolates radiation from the positive internal
mode; center-stable tuning instead selects trajectories of the original
equation that remain near the saddle.  Conditional asymptotic stability and
radiative internal-mode damping of unstable Klein--Gordon solitons have
already been studied \cite{LiLuhrmann2023,BizonRomanczukiewicz2026}, alongside
internal excitations and decay of sphalerons
\cite{MantonRomanczukiewicz2023,NavarroObregonQueiruga2024,NavarroObregonQueiruga2025}.
We compare these two dynamical prescriptions when the leading,
kinematically open FGR coupling is tuned to zero.

We first resolve signed-overlap zeros on a specified finite interval of the
localized branch, including additional thinner-wall roots.  We then examine
two representative points,
$\dlt_\star\simeq0.124166$ and
$\dlt_\dagger\simeq0.227108$, using spatial interference profiles,
half-line scattering, wall deformations, and radial partial differential
equation (PDE) evolution.  These labels denote selected reference points,
not an ordering of all zeros on the branch.  Near $\dlt_\star$, the main
quantitative results are the $A^2$ shift of the radiation minimum, the
$A$-linear width of the region dominated by the third harmonic, and the
distinct higher-order coefficients selected by projected and center-stable
dynamics.  The analysis concerns intrinsic scalar relaxation near a static
critical bubble.  The center-stable construction conditions the unstable
direction and does not stabilize a generic nucleating bubble.

The paper is organized as follows.  Section~\ref{sec:model} defines the
critical-bubble background and localized internal mode.
Section~\ref{sec:perturbation} derives the outgoing-wave hierarchy and the
finite-amplitude continuation of a radiation zero.  Section~\ref{sec:pde}
describes the projected and center-stable radial PDE formulations.
Section~\ref{sec:results} presents the spectral, interference, robustness, and
nonlinear-damping results.  Sections~\ref{sec:discussion} and
\ref{sec:conclusion} give the physical interpretation and conclusions.

\section{Critical bubble and localized internal mode}
\label{sec:model}

We consider a single real scalar order parameter with the effective Lorentzian
Lagrangian density
\[
 \mathcal L=\frac{1}{2}\partial_\mu\phi\,\partial^\mu\phi
 -V(\phi;\dlt).
\]
The temperature dependence of the underlying theory is encoded here in the
effective potential, while the kinetic term is taken in canonical form.
As in Ref.~\cite{InagakiMurakamiLocalized2026}, let
$\phi_{{\rm t},c}$ and $\phi_{{\rm f},c}$ denote the true- and
false-vacuum field values at coexistence, and let $\lambda_c$ be the
corresponding quartic coupling.  We shift the false vacuum to $\phi=0$ and
define
\begin{equation}
 \vev\equiv\phi_{{\rm t},c}-\phi_{{\rm f},c},
 \qquad
 \lambda\equiv\lambda_c.
 \label{eq:reduced-identification}
\end{equation}
We then consider the same reduced finite-temperature free-energy potential
\begin{equation}
 V(\phi;\dlt)
 =\frac{\lambda\vev^4}{4}\,\widehat V(s;\dlt),
 \qquad
 s=\frac{\phi}{\vev},
 \label{eq:physical-potential}
\end{equation}
with
\begin{equation}
 \widehat V(s;\dlt)
 =s^2(1-s)^2-\dlt(3s^2-2s^3).
 \label{eq:dimensionless-potential}
\end{equation}
The false and true vacua remain at $s=0$ and $s=1$, respectively.  Writing
$\Delta V\equiv V(0;\dlt)-V(\vev;\dlt)$ for their free-energy difference,
the normalization of Ref.~\cite{InagakiMurakamiLocalized2026} gives
\begin{equation}
 \Delta V=\frac{\lambda\vev^4}{4}\,\dlt,
 \qquad
 \dlt=\frac{4\,\Delta V}{\lambda\vev^4}.
 \label{eq:delta-definition}
\end{equation}
Thus $\dlt$ parametrizes the vacuum free-energy difference and will be referred to as the reduced supercooling.  No universal map from $\dlt$ to temperature is assumed.  The false vacuum is locally stable for $0<\dlt<1/3$.

The length scale used below is fixed by the planar wall at coexistence.  At
$\dlt=0$, the first integral of the static planar field equation gives
\[
 \frac{1}{2}\left(\frac{\dd\phi}{\dd z}\right)^2=V(\phi;0),
\]
and the wall solution may be written as
\[
 s_{\rm wall}(z)=\frac{1}{2}\left[1-\tanh\left(\frac{z-z_0}{\Lzero}\right)\right].
\]
Substitution into the planar equation fixes the coexistence wall-width
parameter to $\Lzero=2\sqrt{2}/(\vev\sqrt{\lambda})$.  We use this
quantity as the common unit of length and Lorentzian time,
\begin{equation}
 \Lzero=\frac{2\sqrt{2}}{\vev\sqrt{\lambda}},
 \qquad
 \rho=\frac{r}{\Lzero},
 \qquad
 \tau=\frac{t}{\Lzero}.
 \label{eq:units}
\end{equation}
The dimensionless field equation is
\begin{equation}
 \partial_\tau^2 s-\nabla_\rho^2 s+2\widehat V'(s)=0.
 \label{eq:lorentzian-field-equation}
\end{equation}
Indeed, starting from the physical equation
$\partial_t^2\phi-\nabla_r^2\phi+\partial V/\partial\phi=0$ and substituting
$\phi=\vev s$ together with Eq.~\eqref{eq:units} multiplies the potential
term by $\Lzero^2\lambda\vev^2/4=2$, which yields
Eq.~\eqref{eq:lorentzian-field-equation}.

The static $O(3)$ critical bubble obeys
\begin{equation}
 s_b''+\frac{2}{\rho}s_b'=2\widehat V'(s_b)
 =4s_b(1-s_b)(1-2s_b-3\dlt),
 \label{eq:bounce}
\end{equation}
with $s_b'(0)=0$ and $s_b(\infty)=0$.  It is the critical saddle separating initial data that collapse from those that expand.

\begin{figure}[tbp]
 \centering
 \includegraphics[width=0.82\linewidth]{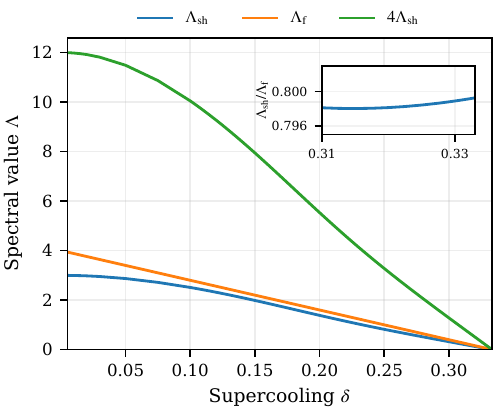}
 \caption{Shape-connected internal-mode eigenvalue $\Lam_{\rm sh}$,
 false-vacuum continuum threshold $\Lam_{\rm f}$, and second-harmonic
 spectral value $4\Lam_{\rm sh}$. The fundamental is bound and the second
 harmonic is open at all resolved samples. The inset displays the binding
 ratio near the spinodal. The curves end at the last sampled point,
 $\dlt=0.333$, where $\Lam_{\rm sh}/\Lam_{\rm f}\simeq0.79926$;
 this termination does not represent a continuum merger.}
 \label{fig:spectral-channels}
\end{figure}

Writing $s=s_b+\eta$, the linear fluctuation operator separates into
partial waves labeled by the angular momentum $\ell=0,1,2,\ldots$.  To
remove the first radial derivative from the fluctuation equation, write a
mode as $\eta=\rho^{-1}u_{\ell m}(\rho)Y_{\ell m}(\Omega)e^{-\ii\omega\tau}$.
Using
$\nabla_\rho^2(\rho^{-1}uY_{\ell m})=\rho^{-1}[u''-\ell(\ell+1)u/\rho^2]Y_{\ell m}$
then reduces the eigenvalue problem to a one-dimensional Schr\"odinger form.
We retain the notation of Ref.~\cite{InagakiMurakamiLocalized2026} and define
the reduced radial Hessian
\begin{equation}
 \mathcal H_\ell
 =-\frac{\dd^2}{\dd\rho^2}
 +\frac{\ell(\ell+1)}{\rho^2}
 +U(\rho;\dlt),
 \label{eq:hessian}
\end{equation}
where
\begin{equation}
 U(\rho;\dlt)
 =2\widehat V''(s_b)
 =4-24s_b+24s_b^2-12\dlt+24\dlt s_b.
 \label{eq:hessian-potential}
\end{equation}
The reduced radial eigenfunctions satisfy
\begin{equation}
 \mathcal H_\ell u_{\ell n}
 =\Lam_{\ell n}u_{\ell n},
 \qquad
 \int_0^\infty u_{\ell n}^2\,\dd\rho=1.
 \label{eq:radial-hessian}
\end{equation}
For the spherically symmetric sector used below,
\begin{equation}
 \mathcal H_0u_-=\Lam_-u_-,
 \qquad
 \mathcal H_0u_{\rm sh}=\Lam_{\rm sh}u_{\rm sh},
 \label{eq:negative-shape-modes}
\end{equation}
where $\Lam_-<0$ is the unstable radial eigenvalue and $u_{\rm sh}$ is the positive $\ell=0$ shape-connected internal mode.  In terms of the branch notation of Ref.~\cite{InagakiMurakamiLocalized2026},
\begin{equation}
 u_{\rm sh}\equiv u_{{\rm sh},0},
 \qquad
 \Lam_{\rm sh}\equiv\Lam_{{\rm sh},0}.
 \label{eq:shape-notation}
\end{equation}
The corresponding dimensionless Lorentzian oscillation frequency is
\begin{equation}
 \omega_{\rm sh}=\sqrt{\Lam_{\rm sh}},
 \qquad
 \omega_{\rm sh}^{\rm phys}=\frac{\sqrt{\Lam_{\rm sh}}}{\Lzero}.
 \label{eq:internal-frequency}
\end{equation}
The false-vacuum continuum begins at
\begin{equation}
 \Lam_{\rm f}=U(\infty)=4-12\dlt.
 \label{eq:false-threshold}
\end{equation}
Across the resolved spectral samples, $\Lam_{\rm sh}<\Lam_{\rm f}$ while
$4\Lam_{\rm sh}>\Lam_{\rm f}$, so the fundamental is bound and the second
harmonic is propagating.  The continuation shown in
Fig.~\ref{fig:spectral-channels} reaches
\begin{equation}
 \dlt=0.333,\qquad
 \Lam_{\rm sh}\simeq0.0031970<\Lam_{\rm f}=0.004.
 \label{eq:resolved-branch-limit}
\end{equation}
This last sampled point is not an inferred spectral endpoint.  It extends the
$\ell=0$ spectral continuation beyond the geometric wall regime analyzed in
Ref.~\cite{InagakiMurakamiLocalized2026}.
Near the spinodal $\dlt=1/3$, the outer domain must grow with the
false-vacuum correlation length.  A fixed box with $\rho_{\max}=25$
produces an apparent threshold crossing near $\dlt=0.33065$; enlarging the
domain instead gives $\Lam_{\rm sh}/\Lam_{\rm f}\simeq0.79897$ there.
The domain and grid checks in Appendix~\ref{app:convergence} exclude
interpreting that finite-box crossing as a mode--continuum merger.

For radial dynamics we use the canonical reduced perturbation
\begin{equation}
 \chi(\rho,\tau)
 =\sqrt{4\pi}\rho\,\eta(\rho,\tau)
 =\sqrt{4\pi}\rho\,[s(\rho,\tau)-s_b(\rho)],
 \label{eq:chi-definition}
\end{equation}
for which the leading internal-mode motion is
\begin{equation}
 \chi_1(\rho,\tau)=u_{\rm sh}(\rho)\cos(\omega_{\rm sh}\tau).
 \label{eq:linear-shape-motion}
\end{equation}

\section{Outgoing-wave perturbation theory}
\label{sec:perturbation}

We remove the unstable radial direction with a fixed projector and derive the
static and second-harmonic responses and their outgoing power.  Expressing the
radiation amplitude as an on-shell continuum overlap gives the radiation-zero
condition.  The same hierarchy yields the higher harmonics and finite-amplitude
corrections required near a zero.

\subsection{Projected perturbative hierarchy and radiated power}

Using the canonical field defined in Eq.~\eqref{eq:chi-definition}, define
\begin{equation}
 \mathcal H_0=-\frac{\dd^2}{\dd\rho^2}+U,
 \qquad
 \langle f,g\rangle=\int_0^\infty f^*(\rho)g(\rho)\,\dd\rho,
 \label{eq:radial-inner-product}
\end{equation}
and the fixed orthogonal projector
\begin{equation}
 \mathsf P_\perp
 =1-|u_-\rangle\langle u_-|.
 \label{eq:fixed-projector}
\end{equation}
Because $u_-$ is an eigenfunction of $\mathcal H_0$, the projector commutes
with the linear operator.  The nonlinear coefficients follow directly from
the Taylor expansion about the bubble,
\[
 2\bigl[\widehat V'(s_b+\eta)-\widehat V'(s_b)\bigr]
 =U\eta+g\eta^2+8\eta^3,
\]
where $g=\widehat V'''(s_b)=12(2s_b-1+\dlt)$ and
$\widehat V''''=24$.  Substituting
$\eta=\chi/(\sqrt{4\pi}\rho)$ and multiplying the radial equation by
$\sqrt{4\pi}\rho$ converts the quadratic and cubic terms into
$g\chi^2/(2\sqrt{\pi}\rho)$ and $2\chi^3/(\pi\rho^2)$, respectively.
Imposing $\chi=\mathsf P_\perp\chi$ therefore gives the constrained
continuum equation
\begin{equation}
 \partial_\tau^2\chi+\mathcal H_0\chi
 +\mathsf P_\perp\left[
 \frac{g(\rho)}{2\sqrt{\pi}\rho}\chi^2
 +\frac{2}{\pi\rho^2}\chi^3
 \right]=0,
 \qquad
 \mathsf P_\perp\chi=\chi,
 \label{eq:projected-chi-equation}
\end{equation}
where
\begin{equation}
 g(\rho)=12\,[2s_b(\rho)-1+\dlt].
 \label{eq:radial-couplings}
\end{equation}
The coupling $g$ changes sign across the wall.  Equation~\eqref{eq:projected-chi-equation}
is equivalently the unprojected equation with a Lagrange-multiplier force,
\begin{equation}
 \begin{aligned}
 \partial_\tau^2\chi+\mathcal H_0\chi
 &+\frac{g}{2\sqrt{\pi}\rho}\chi^2
 +\frac{2}{\pi\rho^2}\chi^3
 =\mu(\tau)u_-,\\
 \mu(\tau)&=\left\langle u_-,
 \frac{g}{2\sqrt{\pi}\rho}\chi^2
 +\frac{2}{\pi\rho^2}\chi^3\right\rangle.
 \end{aligned}
 \label{eq:constraint-force}
\end{equation}
Thus the analytic hierarchy and the real-time PDE use the same fixed
constraint, rather than removing a negative-mode component only after the
perturbative calculation.

Let $A$ denote the small dimensionless amplitude of the localized mode and
write
\begin{equation}
 \begin{aligned}
 \chi&=A\chi_1+A^2\chi_2+A^3\chi_3+\Order(A^4),\\
 \chi_j&=\mathsf P_\perp\chi_j,\\
 \chi_1&=u_{\rm sh}\cos(\omega_{\rm sh}\tau).
 \end{aligned}
 \label{eq:projected-expansion}
\end{equation}
At second order, both a static dressing and a second harmonic are generated:
\begin{equation}
 \chi_2=w_0+
 \operatorname{Re}\left[w_2e^{-2\ii\omega_{\rm sh}\tau}\right].
 \label{eq:second-order-decomposition}
\end{equation}
This follows from
$\chi_1^2=u_{\rm sh}^2\cos^2(\omega_{\rm sh}\tau)
=\tfrac12u_{\rm sh}^2[1+\cos(2\omega_{\rm sh}\tau)]$: the quadratic
nonlinearity therefore drives the zero-frequency and $2\omega_{\rm sh}$
components with the same radial source.  They obey
\begin{equation}
 \mathcal H_0w_0=\mathsf P_\perp \mathcal J_2,
 \qquad
 \left(\mathcal H_0-4\omega_{\rm sh}^2\right)w_2
 =\mathsf P_\perp \mathcal J_2,
 \label{eq:projected-second-order}
\end{equation}
with
\begin{equation}
 \mathcal J_2(\rho)
 =-\frac{g(\rho)u_{\rm sh}^2(\rho)}{4\sqrt{\pi}\rho},
 \qquad
 \langle u_-,w_0\rangle=\langle u_-,w_2\rangle=0.
 \label{eq:w2-source}
\end{equation}
Regularity requires $w_2(0)=0$.  At large radius the outgoing solution has
\begin{equation}
 w_2(\rho)\longrightarrow \mathcal A_2e^{\ii k_2\rho},
 \qquad
 k_2=\sqrt{4\Lam_{\rm sh}-\Lam_{\rm f}}.
 \label{eq:w2-outgoing}
\end{equation}
In practice we impose the finite-radius Robin condition
$w_2'=\ii k_2w_2$ after the bubble potential and source have reached their
asymptotic values.

The reason that the leading second-harmonic result is insensitive to the
constraint can be seen explicitly.  If $w_2^{\rm un}$ denotes the
unprojected solution with source $\mathcal J_2$, then
\begin{equation}
 w_2^{\rm un}=w_2^\perp+c_-u_-,
 \qquad
 c_-=\frac{\langle u_-,\mathcal J_2\rangle}{\Lam_--4\Lam_{\rm sh}},
 \label{eq:w2-decomposition}
\end{equation}
where $w_2^\perp$ is the solution in
Eq.~\eqref{eq:projected-second-order}.  Here and below the superscript
``$\rm un$'' denotes an unprojected quantity, while $\perp$ denotes the
corresponding fixed-projector quantity.  The difference is a localized bound
state and has no outgoing tail.  Consequently
$\mathcal A_2^\perp=\mathcal A_2^{\rm un}$ in the infinite-domain problem,
up to an exponentially small finite-boundary error.

The static response $w_0$ is required for a uniformly valid expansion even
though it does not enter the $3\omega_{\rm sh}$ source.  In a
Poincar\'e--Lindstedt description of the near-periodic core motion on the fast time scale
\cite{Nayfeh1973}, the real part of the order-$A^2$ frequency shift is
\begin{equation}
 \begin{aligned}
 \omega_{\rm sh}(A)
 &=\omega_{\rm sh}+A^2\omega_{\rm sh}^{(2)}+\Order(A^4),\\
 \omega_{\rm sh}^{(2)}
 &=\frac{1}{2\omega_{\rm sh}}\operatorname{Re}
 \left\langle u_{\rm sh},\,
 \frac{g u_{\rm sh}w_0}{\sqrt{\pi}\rho}
 +\frac{g u_{\rm sh}w_2^\perp}{2\sqrt{\pi}\rho}\right.\\[-1mm]
 &\hspace{5.5em}\left.
 +\frac{3u_{\rm sh}^3}{2\pi\rho^2}
 \right\rangle.
 \end{aligned}
 \label{eq:projected-frequency-shift}
\end{equation}
The imaginary on-shell part is more transparently represented by the
radiated power and the slow amplitude equation derived next.

The outgoing harmonic determines the radiated power and the slow evolution
of the internal-mode amplitude.

Appendix~\ref{app:flux} derives the dimensionless cycle-averaged flux of an
outgoing reduced spherical wave.  For the second harmonic,
\begin{equation}
 \Pbar_2=\Gamma_2A^4,
 \qquad
 \Gamma_2=\frac{1}{2}(2\omega_{\rm sh})k_2|\mathcal A_2|^2.
 \label{eq:gamma2}
\end{equation}
The leading internal-mode energy is
\begin{equation}
 E_{\rm sh}=\frac{1}{2}\Lam_{\rm sh}A^2.
 \label{eq:mode-energy}
\end{equation}
With $\int u_{\rm sh}^2\dd\rho=1$ and
$\mathcal H_0u_{\rm sh}=\Lam_{\rm sh}u_{\rm sh}$, this is the quadratic
energy $\tfrac12\langle\chi_\tau,\chi_\tau\rangle+
\tfrac12\langle\chi,\mathcal H_0\chi\rangle$ evaluated on the linear
oscillation.  Energy balance $\dd E_{\rm sh}/\dd\tau=-\Pbar_2$ gives
\begin{equation}
 \frac{\dd A}{\dd\tau}
 =-\frac{\Gamma_2}{\Lam_{\rm sh}}A^3,
 \label{eq:amplitude-equation-second}
\end{equation}
and therefore
\begin{equation}
 A(\tau)
 =\frac{A_0}{\sqrt{1+2(\Gamma_2/\Lam_{\rm sh})A_0^2\tau}}.
 \label{eq:ordinary-decay}
\end{equation}
Here $A_0\equiv A(0)$ is the initial internal-mode amplitude.  As shown in
Eq.~\eqref{eq:fgr-overlap} below, the outgoing amplitude is proportional to
the on-shell quadratic FGR overlap.  Therefore $\mathcal F_2\neq0$ implies
$\mathcal A_2\neq0$ and hence $\Gamma_2>0$ in
Eq.~\eqref{eq:gamma2}; Eq.~\eqref{eq:ordinary-decay} then gives
$A\sim\tau^{-1/2}$.

\subsection{Radiation zero and higher-harmonic hierarchy}

Let $u_k^{\rm reg}$ be a real solution of the homogeneous continuum equation at $k=k_2$, regular at the origin.  The outgoing Green-function representation expresses the radiation amplitude as a nonzero normalization factor multiplying the source overlap with this regular solution.  Hence the zeros of the outgoing amplitude are controlled by
\begin{equation}
 \mathcal F_2(\dlt)
 =\langle u_k^{\rm reg},\mathsf P_\perp \mathcal J_2\rangle
 =\int_0^\infty u_k^{\rm reg}(\rho)\mathcal J_2(\rho)\,\dd\rho.
 \label{eq:fgr-overlap}
\end{equation}
The second equality follows from the orthogonality of a continuum
eigenfunction to the negative bound state.  A zero of $\mathcal F_2$ removes
the leading outgoing second harmonic even though $k_2$ remains real; its
location is therefore unchanged by the fixed projection.

At the selected root $\dlt_\star$ studied below, the third harmonic is
nonzero and becomes asymptotically leading.  With
\begin{equation}
 \chi^{(3)}_{3\omega_{\rm sh}}
 =A^3\operatorname{Re}\left[w_3(\rho)e^{-3\ii\omega_{\rm sh}\tau}\right],
\end{equation}
the consistently projected third-order equation is
\begin{equation}
 \left[\mathcal H_0-9\omega_{\rm sh}^2\right]w_3
 =\mathsf P_\perp \mathcal J_3,
 \qquad
 \langle u_-,w_3\rangle=0,
 \label{eq:w3-equation}
\end{equation}
where
\begin{equation}
 \mathcal J_3
 =-\frac{g u_{\rm sh}w_2^\perp}{2\sqrt{\pi}\rho}
  -\frac{u_{\rm sh}^3}{2\pi\rho^2}.
 \label{eq:projected-w3-source}
\end{equation}
The static dressing $w_0$ contributes to the fundamental harmonic and hence
to Eq.~\eqref{eq:projected-frequency-shift}, but not to the $3\omega_{\rm sh}$
component.  By contrast, replacing $w_2^{\rm un}$ with $w_2^\perp$ changes
the radiative third-order source:
\begin{equation}
 \mathcal J_3^\perp
 =\mathsf P_\perp\left[
 \mathcal J_3^{\rm un}+c_-\frac{g}{2\sqrt{\pi}\rho}u_{\rm sh}u_-\right].
 \label{eq:third-source-correction}
\end{equation}
This is the term missed if only the final $w_3$ equation is projected.
If $w_3\to\mathcal A_3e^{\ii k_3\rho}$, where $k_3=\sqrt{9\Lam_{\rm sh}-\Lam_{\rm f}}$, then
\begin{equation}
 \Pbar_3=\Gamma_3^\perp A^6,
 \qquad
 \Gamma_3^\perp=\frac{1}{2}(3\omega_{\rm sh})k_3|\mathcal A_3|^2.
 \label{eq:gamma3}
\end{equation}
Using the same quadratic mode energy as in Eq.~\eqref{eq:mode-energy},
$\dd E_{\rm sh}/\dd\tau=-\Pbar_3$ first gives
\[
 \frac{\dd A}{\dd\tau}
 =-\frac{\Gamma_3^\perp}{\Lam_{\rm sh}}A^5.
\]
Integrating this equation yields
\begin{equation}
 A(\tau)
 =\frac{A_0}{[1+4(\Gamma_3^\perp/\Lam_{\rm sh})A_0^4\tau]^{1/4}},
 \label{eq:zero-decay}
\end{equation}
which gives $A\sim\tau^{-1/4}$.

\subsection{Finite-amplitude continuation}
\label{subsec:fourth-order-theory}

In this subsection $w_j$ denotes the fixed-projector response unless otherwise stated.

The displacement of the finite-amplitude minimum can be predicted without
fitting the PDE valley.  Introduce the Poincar\'e--Lindstedt phase
\begin{equation}
 \theta=\omega_{\rm sh}(A)\tau,
 \qquad
 \omega_{\rm sh}(A)=\omega_{\rm sh}+A^2\omega_{\rm sh}^{(2)}+\Order(A^4),
\end{equation}
and extend the expansion through the cubic responses,
\begin{align}
 \chi={}&A u_{\rm sh}\cos\theta
 +A^2\left[w_0+\operatorname{Re}(w_2e^{-2\ii\theta})\right]\nonumber\\
 &+A^3\operatorname{Re}\left(w_1e^{-\ii\theta}+w_3e^{-3\ii\theta}\right)
 +\Order(A^4).
 \label{eq:pl-through-cubic}
\end{align}
The nonresonant fundamental correction is chosen with
$\langle u_{\rm sh},w_1\rangle=0$.  In the notation of
Eq.~\eqref{eq:radial-couplings}, the cubic sources can be written directly as
\begin{align}
 \mathcal J_1^{(3)}={}&
 2\omega_{\rm sh}\omega_{\rm sh}^{(2)}u_{\rm sh}
 -\frac{g u_{\rm sh}w_0}{\sqrt{\pi}\rho}
 -\frac{g u_{\rm sh}w_2}{2\sqrt{\pi}\rho}
 -\frac{3u_{\rm sh}^3}{2\pi\rho^2},\\
 \mathcal J_3^{(3)}={}&
 -\frac{g u_{\rm sh}w_2}{2\sqrt{\pi}\rho}
 -\frac{u_{\rm sh}^3}{2\pi\rho^2}.
\end{align}
At order $A^3$ in the fundamental harmonic, the nonresonant correction
$w_1$ satisfies
\[
 (\mathcal H_0-\omega_{\rm sh}^2)w_1
 =\mathsf P_\perp\mathcal J_1^{(3)},
 \qquad
 \langle u_{\rm sh},w_1\rangle=0.
\]
On the negative-mode-projected radial subspace,
$\mathcal H_0-\omega_{\rm sh}^2$ is self-adjoint and has
$u_{\rm sh}$ in its kernel.  The Fredholm solvability condition therefore
requires the fundamental source to be orthogonal to $u_{\rm sh}$.  Using
$\langle u_{\rm sh},u_{\rm sh}\rangle=1$ gives
\begin{align*}
 0&=\operatorname{Re}\langle u_{\rm sh},\mathcal J_1^{(3)}\rangle\\
 &=2\omega_{\rm sh}\omega_{\rm sh}^{(2)}
 -\operatorname{Re}\left\langle u_{\rm sh},
 \frac{g u_{\rm sh}w_0}{\sqrt{\pi}\rho}
 +\frac{g u_{\rm sh}w_2}{2\sqrt{\pi}\rho}\right.\\[-1mm]
 &\hspace{7em}\left.
 +\frac{3u_{\rm sh}^3}{2\pi\rho^2}
 \right\rangle .
\end{align*}
Solving this condition for $\omega_{\rm sh}^{(2)}$ gives precisely
Eq.~\eqref{eq:projected-frequency-shift}.

The fourth-order second-harmonic problem is a separate step.  Collecting the
terms oscillating at $2\omega_{\rm sh}$ gives
\begin{equation}
 (\mathcal H_0-4\omega_{\rm sh}^2)w_2^{(4)}
 =\mathsf P_\perp\mathcal J_2^{(4)},
 \label{eq:w2-fourth-order}
\end{equation}
where
\begin{align}
 \mathcal J_2^{(4)}={}&
 8\omega_{\rm sh}\omega_{\rm sh}^{(2)}w_2
 -\frac{g}{2\sqrt{\pi}\rho}
 \left[2w_0w_2+u_{\rm sh}(w_1+w_3)\right]\nonumber\\
 &-\frac{3u_{\rm sh}^2}{\pi\rho^2}(w_0+w_2).
 \label{eq:source2-fourth-order}
\end{align}
The four terms have distinct origins: the first term in
Eq.~\eqref{eq:source2-fourth-order} comes from expanding the
$2\omega_{\rm sh}$ operator with the amplitude-dependent frequency; the
quadratic nonlinearity produces the products $2w_0w_2$ and
$u_{\rm sh}(w_1+w_3)$; and the cubic nonlinearity produces the
$u_{\rm sh}^2(w_0+w_2)$ contribution.  Thus the frequency correction,
static and second-harmonic dressing, nonresonant fundamental response, and
outgoing third harmonic are all needed for the leading finite-amplitude
shift.

Normalize the regular continuum solution by
$u_{k_2}^{\rm reg}(0)=0$ and $(u_{k_2}^{\rm reg})'(0)=1$.  In the notation
below, the superscripts $(2)$ and $(4)$ label the $O(A^2)$ and $O(A^4)$
second-harmonic source contributions, respectively.  We then define
\begin{align}
 \mathcal F_2^{(2)}(\dlt)
 &=\langle u_{k_2}^{\rm reg},\mathcal J_2\rangle,\\
 \mathcal F_2^{(4),\perp}(\dlt)
 &=\langle u_{k_2}^{\rm reg},\mathsf P_\perp\mathcal J_2^{(4)}\rangle.
 \label{eq:fourth-order-overlaps}
\end{align}
Let $\dlt_\star$ denote the selected root near $0.124166$ of
$\mathcal F_2^{(2)}(\dlt)=0$.  Near the root, the total second-harmonic overlap entering the outgoing amplitude has the local expansion
\[
 \mathcal F_{2,\rm tot}(\dlt,A)
 =A^2\left[\mathcal F_2^{(2)}(\dlt)
 +A^2\mathcal F_2^{(4),\perp}(\dlt)+\Order(A^4)\right].
\]
Setting $\dlt=\dlt_\star+c_2A^2$ and using
$\mathcal F_2^{(2)}(\dlt_\star)=0$ gives, at the first nonvanishing
order,
\[
 \mathcal F_{2,\rm tot}=A^4\left[c_2\,\partial_\dlt\mathcal F_2^{(2)}
 +\mathcal F_2^{(4),\perp}\right]+\Order(A^6).
\]
The radiated power is proportional to $|\mathcal F_{2,\rm tot}|^2$, so minimizing this
quantity with respect to the real displacement $c_2$ gives
\begin{equation}
 \begin{aligned}
 c_2^\perp&=-\frac{\operatorname{Re}\left[
 (\partial_\dlt\mathcal F_2^{(2)})^*\mathcal F_2^{(4),\perp}\right]}
 {|\partial_\dlt\mathcal F_2^{(2)}|^2},\\
 \dlt_{\rm min}^\perp(A)&=\dlt_\star+c_2^\perp A^2+\Order(A^4).
 \end{aligned}
 \label{eq:fourth-order-zero-shift}
\end{equation}
Removing $\mathsf P_\perp$ throughout gives the corresponding center-stable
coefficient $c_2^{\rm CS}$; here and below the superscript ${\rm CS}$ denotes
a center-stable quantity.  The two need not agree because the localized
negative-mode component of $w_2$ re-enters the fourth-order source.

\section{Real-time radial PDE formulations}
\label{sec:pde}

We use two real-time radial formulations.  Fixed-projector dynamics isolates
radiation damping of the positive internal mode.  Center-stable shooting
instead retains the full saddle dynamics while removing the exponentially
growing negative-mode solution.  Both formulations test the perturbative
radiation coefficients and their finite-amplitude continuation.

\subsection{Projected evolution and negative-mode control}

The real-time calculation evolves the same canonical field and fixed
projector as Eqs.~\eqref{eq:chi-definition}--\eqref{eq:projected-chi-equation}.
After adding a numerical sponge, the continuum equation represented by the
projected time step is
\begin{equation}
 \chi_{\tau\tau}+\mathcal H_0\chi
 +\mathsf P_\perp\left[
 \frac{g(\rho)}{2\sqrt{\pi}\rho}\chi^2
 +\frac{2}{\pi\rho^2}\chi^3
 +\sigma(\rho)\chi_\tau\right]=0.
 \label{eq:chi-pde}
\end{equation}
Here $\sigma(\rho)$ is the absorbing sponge profile; it is supported only
near the outer boundary and removes outgoing waves before they reflect back
into the interaction region.  Regularity gives $\chi(0,\tau)=0$.

The constrained evolution enforces
\begin{equation}
 \langle u_-,\chi\rangle
 =\langle u_-,\chi_\tau\rangle=0.
 \label{eq:constraint}
\end{equation}
At each time step the conservative acceleration is replaced by
\begin{equation}
 \chi_{\tau\tau}\longrightarrow
 \chi_{\tau\tau}
 -u_-\langle u_-,\chi_{\tau\tau}\rangle,
 \label{eq:acceleration-projection}
\end{equation}
and both the field and velocity are reprojected after the update.  This
defines the codimension-one projected dynamics in which expansion and collapse
are removed so that radiation damping of the positive internal mode can be
measured.

The initial undressed amplitude scans use
\begin{equation}
 \chi(\rho,0)=A u_{\rm sh}(\rho),
 \qquad
 \chi_\tau(\rho,0)=0,
 \label{eq:pde-initial-data}
\end{equation}
while precision third-harmonic extractions use the consistently projected
second-order dressing
\begin{equation}
 \begin{aligned}
 \chi(\rho,0)&=A u_{\rm sh}
 +A^2\left[w_0+\operatorname{Re}w_2^\perp\right],\\
 \chi_\tau(\rho,0)&=2A^2\omega_{\rm sh}\operatorname{Im}w_2^\perp.
 \end{aligned}
 \label{eq:dressed-pde-initial-data-method}
\end{equation}
The velocity follows by differentiating
$\operatorname{Re}[w_2e^{-2\ii\omega_{\rm sh}\tau}]$ at $\tau=0$; hence the
imaginary part of the outgoing dressing fixes the initial $O(A^2)$ velocity.
The production evolution uses a projected second-order velocity--Verlet
update.  Unless stated otherwise, the radial domain is $0\le\rho\le60$, the
detector is at $\rho=20$, and the sponge occupies $45<\rho<60$.  Here $\Delta\rho$ and $\Delta\tau$ denote the radial and temporal grid
spacings.  The baseline grid uses $\Delta\rho=0.04$ and
$\Delta\tau=0.012$; precision calculations reach $\Delta\rho=0.01$ and
$\Delta\tau=0.003$.  Harmonic amplitudes are
obtained from simultaneous fits of the fundamental through fourth harmonics
with Hann or Blackman windows at fixed retarded time.

\subsection{Unconstrained center-stable shooting}
\label{subsec:center-stable-method}

The fixed projector isolates slow radiation but changes the localized
second-order response entering higher harmonics.  We therefore also evolve
the original unconstrained radial equation.  The initial data are dressed
with the unprojected responses,
\begin{align}
 \chi(\rho,0)
 &=A u_{\rm sh}+A^2\left[w_0^{\rm un}+\operatorname{Re}w_2^{\rm un}\right]
 +\alpha u_-,\\
 \chi_\tau(\rho,0)
 &=2A^2\omega_{\rm sh}(A)\operatorname{Im}w_2^{\rm un}
 +\kappa\alpha u_-,
 \qquad \kappa=\sqrt{-\Lam_-}.
 \label{eq:cs-initial-data}
\end{align}
The final terms vary the initial state only along the growing linear
phase-space direction.  To see the shooting condition explicitly, write the
negative-mode coordinate in the linear regime as
\[
 x_-(\tau)=x_-^{(2)}(\tau)+C_+e^{\kappa\tau}+C_-e^{-\kappa\tau},
 \qquad x_-=\langle u_-,\chi\rangle.
\]
Then
$\dot x_-+\kappa x_--(\dot x_-^{(2)}+\kappa x_-^{(2)})
=2\kappa C_+e^{\kappa\tau}$, so requiring this combination to vanish removes
only the growing homogeneous component.  The shooting parameter $\alpha$ is
therefore tuned so that at a terminal time $T$
\begin{equation}
 [\dot x_-+\kappa x_-]_{\tau=T}
 =[\dot x_-^{(2)}+\kappa x_-^{(2)}]_{\tau=T},
 \label{eq:cs-terminal}
\end{equation}
where $x_-^{(2)}$ is the known second-order negative-mode response.  This
removes the growing homogeneous solution while leaving the decaying
component free.  The state and tangent with respect to $\alpha$ are evolved
together and Newton continuation is carried to $T=46$.  A terminal-time check at
$T=40,46,52$, with fixed $\Delta\rho=0.02$, $\Delta\tau=0.006$,
and $A=0.06,0.08,0.10$, changes the tuned $\alpha_*(A)$ by less than
$3.4\times10^{-10}$ relative to its $T=46$ value.
This tests the shooting horizon at fixed discretization, separately from
the grid and amplitude extrapolations below; numerical details are given in
Appendix~\ref{app:convergence}.

\section{Results}
\label{sec:results}

We first identify the radiation zeros and resolve their spatial interference
structure and robustness.  We then compare perturbation theory with projected
and center-stable PDE dynamics, including the finite-amplitude displacement of
the radiation minimum and the resulting slow-envelope evolution.

Representative values are rounded for display; derived coefficients,
relative differences, and fit diagnostics use unrounded data.  Extra digits
are retained where needed for convergence checks or simulation inputs.

\subsection{Radiation zeros, interference, and robustness}

Across the zero neighborhoods considered here, the second harmonic remains
propagating while the signed on-shell source overlap changes sign.  Its
radiation amplitude therefore vanishes by open-channel cancellation rather
than by a kinematic threshold effect.  Supercooling tunes the radiation form
factor of the extended wall.

\begin{figure*}[t]
 \centering
 \includegraphics[width=0.92\textwidth]{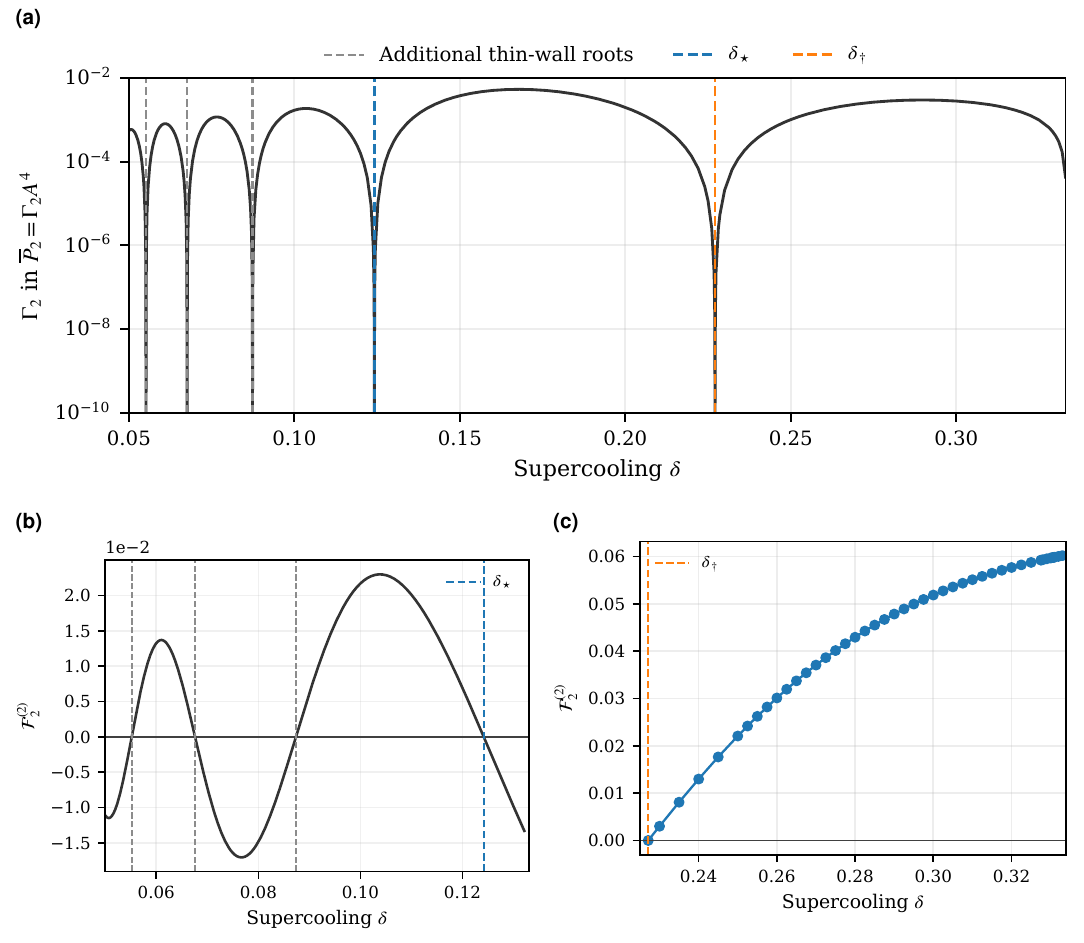}
 \caption{Power and signed-overlap diagnostics of open-channel radiation zeros.
 (a) Second-harmonic power on the declared interval; gray dashed lines mark
 the three additional thinner-wall roots, while blue and orange lines identify
 $\dlt_\star$ and $\dlt_\dagger$. (b) Signed thin-wall overlap, resolving the
 three additional sign changes and the crossing at $\dlt_\star$.
 (c) Signed samples beyond $\dlt_\dagger$, with enlarged outer domains toward
 the spinodal. The refined scan uses the original resolved samples below $\dlt=0.25$
 and a recomputed thick-wall tail through $\dlt=0.333$, with outer domains
 scaled to the false-vacuum correlation length.  The plotted floor in
 panel (a) is for visibility, not a nonzero residual power at the roots.
 The figure makes no claim about a complete zero count as $\dlt\to0$.}
 \label{fig:gamma2-scan}
\end{figure*}

Figure~\ref{fig:gamma2-scan} displays the sampled interval
$0.05\le\dlt\le0.333$.  Resolving the signed overlap is essential: a
coarse scan of the nonnegative coefficient $\Gamma_2$ can miss narrow
zeros on the thinner-wall side.  The refined calculation finds three roots
near $0.055176$, $0.067599$, and $0.087352$, in addition to the
representative points $\dlt_\star$ and $\dlt_\dagger$.
Their grid convergence is given in Table~\ref{tab:zero-convergence}.
We do not infer an exhaustive zero count over the full existence range
$0<\dlt<1/3$:
in particular, the interval $0<\dlt<0.05$ is not covered by this resolved
scan.  The oscillatory thin-wall overlap motivates sampling in inverse
supercooling, which more directly resolves the changing wall radius.

Beyond $\dlt_\dagger$, $\Gamma_2$ reaches a broad maximum near
$\dlt\simeq0.29$ before decreasing along the resolved thick-wall tail.
The signed-overlap samples remain positive for
$\dlt_\dagger<\dlt\le0.333$, reaching $0.06017$ at the last point.
This resolved tail does not show another sign reversal; it does not certify
the absence of zeros between all samples or beyond the resolved interval.
Representative outgoing-wave values appear in
Table~\ref{tab:radiation-scan}.  An ordinary comparison point is $\dlt=0.1$,
where, using the same dataset and displayed precision as
Table~\ref{tab:radiation-scan},
\begin{equation}
 \Lam_{\rm sh}=2.5113,
 \qquad
 \Gamma_2=1.6822\times10^{-3}.
 \label{eq:reference-benchmark}
\end{equation}

\begin{table*}[t]
 \centering
 \caption{Representative outgoing-wave results selected from Fig.~\ref{fig:gamma2-scan}.  The entries sample the thin-wall side, neighborhoods of the representative zeros, the broad intermediate maximum, and the resolved thick-wall tail. All entries are recomputed at nominal $h=0.0025$, with $\rho_{\max}=\max(40,30/\sqrt{\Lam_{\rm f}})$. Computed quantities are rounded to five significant figures; this is not a continuum error bound.}
 \label{tab:radiation-scan}
 \begin{tabular}{ccccc}
  \toprule
  $\dlt$ & $\Lam_{\rm sh}$ & $\omega_{\rm sh}$ & $k_2$ & $\Gamma_2$ \\
  \midrule
  0.075 & 2.7162 & 1.6481 & 2.7865 & $1.0857\times10^{-3}$ \\
  0.1 & 2.5113 & 1.5847 & 2.6917 & $1.6822\times10^{-3}$ \\
  0.12 & 2.3169 & 1.5221 & 2.5899 & $1.5532\times10^{-4}$ \\
  0.13 & 2.2111 & 1.4870 & 2.5307 & $2.9992\times10^{-4}$ \\
  0.17 & 1.7469 & 1.3217 & 2.2422 & $5.2721\times10^{-3}$ \\
  0.22 & 1.1480 & 1.0714 & 1.7978 & $1.3499\times10^{-4}$ \\
  0.225 & 1.0911 & 1.0446 & 1.7506 & $1.1509\times10^{-5}$ \\
  0.227 & 1.0686 & 1.0337 & 1.7316 & $2.9605\times10^{-8}$ \\
  0.25 & 0.81877 & 0.90486 & 1.5083 & $1.0026\times10^{-3}$ \\
  0.29 & 0.41636 & 0.64526 & 1.0703 & $2.9492\times10^{-3}$ \\
  0.32 & 0.12770 & 0.35735 & 0.59229 & $1.4881\times10^{-3}$ \\
  0.3305 & 0.027164 & 0.16482 & 0.27324 & $3.5298\times10^{-4}$ \\
  0.333 & 0.0031970 & 0.056542 & 0.093745 & $4.2488\times10^{-5}$ \\
  \bottomrule
 \end{tabular}
\end{table*}

Second-order Richardson extrapolation gives the reference points
(full values are retained in Table~\ref{tab:zero-convergence})
\begin{equation}
 \dlt_\star\simeq0.124166,
 \qquad
 \dlt_\dagger\simeq0.227108.
 \label{eq:two-zeros}
\end{equation}
Table~\ref{tab:zero-convergence} also gives the three additional thinner-wall
roots, denoted $\dlt_a$, $\dlt_b$, and $\dlt_c$ within the displayed
interval.  The ratios of successive grid corrections are approximately four,
as expected for the centered second-order Hessian discretization.  The
associated open-channel wave numbers are approximately $2.8346$, $2.8074$,
and $2.7443$.  A separate calculation with $\rho_{\max}=40$ reproduces
these three roots at nominal $h=0.005$ within $1.3\times10^{-11}$;
the nominal-grid convention is explained in Appendix~\ref{app:protocol}.

The detailed deformation and PDE tests below concern the two
representative roots $\dlt_\star$ and $\dlt_\dagger$.  The former remains
the focus of the finite-amplitude and center-stable analysis.  The latter
provides a comparison deeper in the thick-wall regime.  The three
additional roots are established by signed overlaps, grid convergence,
and the enlarged-domain check.  They also pass the independent outgoing-BVP
and Jost checks in Table~\ref{tab:jost-additional-zeros}; their higher-order
dynamics are not evaluated.

\begin{table*}[t]
 \centering
 \caption{Grid convergence of five resolved radiation zeros.  The column
 headings give nominal Hessian spacings; the final column extrapolates the
 two finest grids.  Extra digits document convergence rather than an
 uncertainty bound on the continuum roots.}
 \label{tab:zero-convergence}
 \small
 \begin{tabular}{ccccc}
  \toprule
  root & $h=0.0100$ & $h=0.0050$ & $h=0.0025$ & $h\to0$ \\
  \midrule
  $\dlt_a$ & 0.0551756591 & 0.0551760034 & 0.0551760895 & 0.0551761182 \\
  $\dlt_b$ & 0.0675982594 & 0.0675986603 & 0.0675987606 & 0.0675987940 \\
  $\dlt_c$ & 0.0873515611 & 0.0873520279 & 0.0873521446 & 0.0873521835 \\
  $\dlt_\star$ & 0.1241651320 & 0.1241655983 & 0.1241657149 & 0.1241657537 \\
  $\dlt_\dagger$ & 0.2271084733 & 0.2271078766 & 0.2271077275 & 0.2271076778 \\
  \bottomrule
 \end{tabular}
\end{table*}

\begin{figure*}[t]
 \centering
 \includegraphics[width=\textwidth]{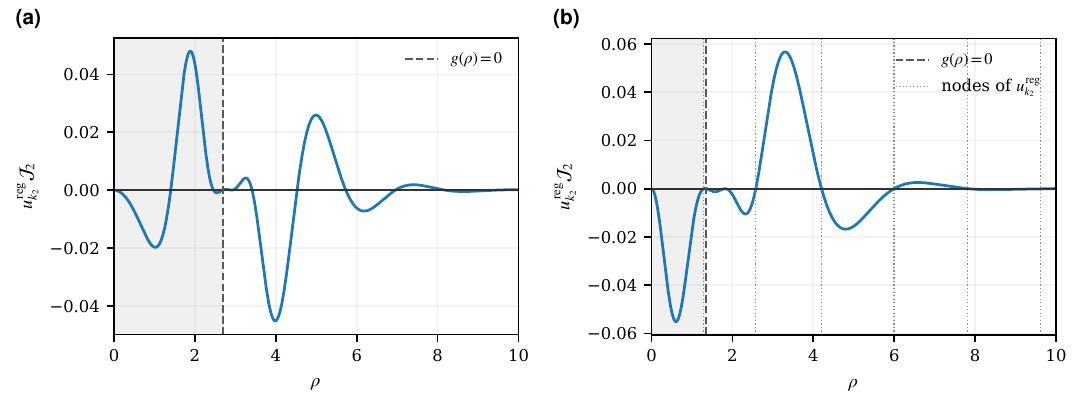}
 \caption{Local weighted on-shell quadratic source at the two representative
 radiation zeros. (a) At $\dlt_\star$, the dashed line marks
 $g(\rho_g)=0$ and the shaded region is the inner side $\rho<\rho_g$;
 the cancellation is organized mainly by the inner--outer structure of the
 curved wall. (b) At $\dlt_\dagger$, the dashed line again marks
 $g(\rho_g)=0$, while dotted lines mark nodes of $u_{k_2}^{\rm reg}$;
 the profile contains several continuum-phase lobes in the outer region.}
 \label{fig:interference-local}
\end{figure*}

At $\dlt_\star$,
\begin{equation}
 \Lam_{\rm sh}=2.2734,
 \quad
 \Lam_{\rm f}=2.5100,
 \quad
 k_2=2.5659.
 \label{eq:reference-star-spectrum}
\end{equation}
The channel is therefore not near closure.  Instead, the source-weighted
overlap cancels.  Let $\rho_g$ denote the radius at which the quadratic
coupling changes sign, $g(\rho_g)=0$; at $\dlt_\star$,
$\rho_g=2.6885$.  Splitting Eq.~\eqref{eq:fgr-overlap} there gives
\begin{equation}
 \mathcal F_2^{\rm inner}=0.01364,
 \qquad
 \mathcal F_2^{\rm outer}=-0.01364.
 \label{eq:inner-outer-cancellation}
\end{equation}
To separate the local phase structure from the accumulated cancellation, we
show these two diagnostics in separate figures.  Figure~\ref{fig:interference-local}
displays only the local weighted source $u_{k_2}^{\rm reg}\mathcal J_2$.  The shaded
region is $\rho<\rho_g$, the dashed line marks $g(\rho_g)=0$, and, for
$\dlt_\dagger$, the dotted lines mark nodes of the regular continuum solution
$u_{k_2}^{\rm reg}$.  At $\dlt_\star$ the dominant sign organization is set by the
wall-scale inner--outer partition.  At $\dlt_\dagger$ the regular continuum
solution already has a node at $\rho\simeq1.2936$, just inside
$\rho_g=1.3563$, followed by nodes near $2.5788$, $4.2031$, $5.9938$,
$7.8067$, and $9.6218$.  These nodes partition the outer region into
alternating phase intervals.  Thus the sign changes of
$u_{k_2}^{\rm reg}\mathcal J_2$ at $\dlt_\dagger$ reflect both the sign reversal of
the quadratic source through $g(\rho)$ and the accumulated continuum phase.

To compare neighborhoods of the two roots, define the root-centered detuning
$\Delta\dlt_j\equiv\dlt-\dlt_j$ for $j=\star,\dagger$.  On the
$h=0.0025$ grid we use $\dlt_\dagger=0.2271077275$ and evaluate
$\mathcal I(R)=\int_0^Ru_{k_2}^{\rm reg}\mathcal J_2\,\dd\rho$ at symmetric
detunings $\Delta\dlt_\dagger=\pm2\times10^{-4}$.  The far-radius values are
\begin{equation}
 \mathcal I(\infty)=
 \begin{cases}
 -2.0908\times10^{-4}, & \Delta\dlt_\dagger=-2\times10^{-4},\\
 -7.99\times10^{-14}, & \Delta\dlt_\dagger=0,\\
 +2.0891\times10^{-4}, & \Delta\dlt_\dagger=+2\times10^{-4}.
 \end{cases}
 \label{eq:reference-dagger-overlap-sign}
\end{equation}
The sign reversal agrees with the independently measured positive slope
$\partial_\dlt\mathcal F_2^{(2)}\simeq1.0450$.
At the central root, splitting the total overlap at $\rho_g=1.3563$ gives
\begin{equation}
 \mathcal F_{2,\rm in}^{(2)}=-0.03429,
 \qquad
 \mathcal F_{2,\rm out}^{(2)}=+0.03429,
 \label{eq:reference-dagger-inner-outer}
\end{equation}
so the total inner--outer cancellation remains exact.  The outer term is not
sign definite: the first four outer phase intervals contribute
$-0.004568$, $+0.05398$, $-0.01750$, and $+0.002735$, respectively.

\begin{figure*}[t]
 \centering
 \includegraphics[width=\textwidth]{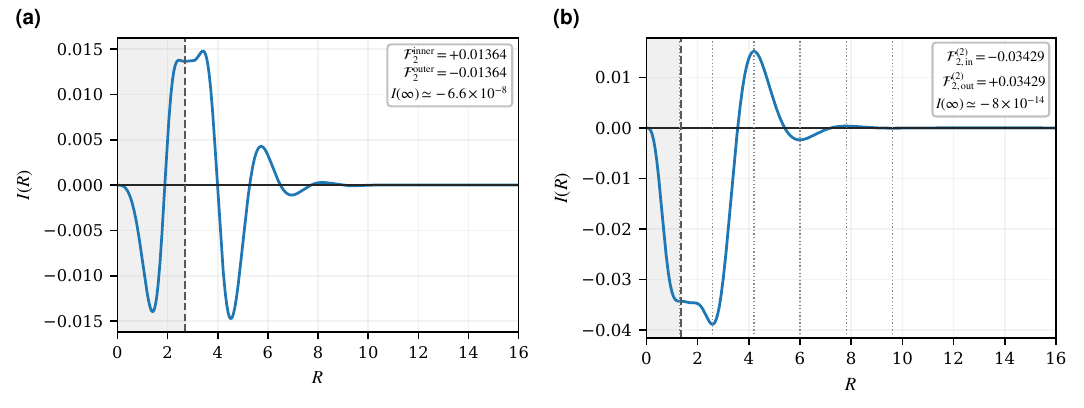}
 \caption{Cumulative overlap
 $I(R)=\int_0^Ru_{k_2}^{\rm reg}\mathcal J_2\,\dd\rho$ at the two
 representative radiation zeros. (a) Cumulative inner--outer cancellation
 at $\dlt_\star$. (b) Cumulative sum of alternating outer phase lobes at
 $\dlt_\dagger$. The dashed line marks $g(\rho_g)=0$; dotted lines in
 panel (b) mark nodes of $u_{k_2}^{\rm reg}$. The boxed annotations give the
 inner/outer decomposition and far-radius total. The small nonzero terminal
 value in panel (a), $I(\infty)\simeq-6.6\times10^{-8}$, is a
 finite-grid/root-location residual of the plotted calculation; it does not
 shift the extrapolated root in Table~\ref{tab:zero-convergence}.}
 \label{fig:interference-cumulative}
\end{figure*}

Figure~\ref{fig:interference-cumulative} isolates the cumulative integral
from the local weighted source.  In both panels the heavier horizontal line is
$I=0$.  The numerical inner/outer contributions and far-radius totals are
listed in boxes in the large-$R$ region, leaving the data curves unobstructed.
For $\dlt_\star$, the cumulative overlap is naturally summarized by the
single inner and outer contributions in Eq.~\eqref{eq:inner-outer-cancellation}.
For $\dlt_\dagger$, the node positions partition the outer region into
successive phase intervals, and the cumulative curve records their alternating
addition before returning to zero.

\begin{figure*}[t]
 \centering
 \includegraphics[width=\textwidth]{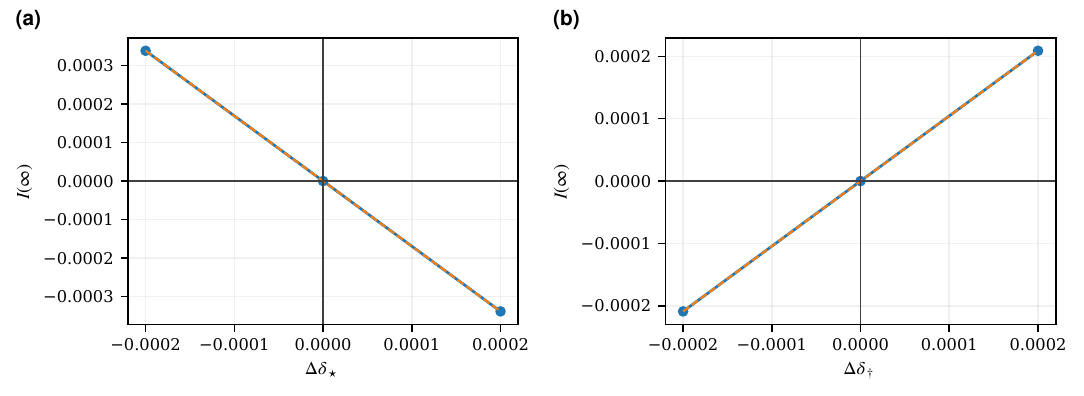}
 \caption{Far-radius cumulative-overlap sign reversal at symmetric detunings
 around the two representative radiation zeros. The detuned overlap
 $I(\infty)$ crosses through zero with (a) negative slope at $\dlt_\star$
 and (b) positive slope at $\dlt_\dagger$, directly displaying the
 simple-zero character of each open-channel cancellation.}
 \label{fig:detuning-sign-reversal}
\end{figure*}

For $\dlt_\star$ and $\dlt_\dagger$, we also examine the far-radius overlap
as a function of detuning.  A simple zero must cross the horizontal axis
linearly, so this diagnostic tests the sign reversal separately from the
spatial accumulation shown in Fig.~\ref{fig:interference-cumulative}.
Figure~\ref{fig:detuning-sign-reversal} shows this behavior for symmetric
root-centered detunings $\Delta\dlt_j=\pm2\times10^{-4}$.

\label{subsec:jost-boundary-check}

To exclude a finite-box origin for the cancellation, define
\begin{equation}
 \mathcal L_\omega=-\frac{\dd^2}{\dd\rho^2}+U(\rho)-\omega^2.
\end{equation}
Let $\varphi_\alpha$ be the regular continuum solution with
$\varphi_\alpha(0)=0$, $\varphi_\alpha'(0)=\alpha$, and let the outgoing
Jost solution be normalized by $f_C^{(+)}\to C e^{\ii k\rho}$.  The retarded
radial Green function is
\begin{equation}
 G^+(\rho,\rho')=-\frac{\varphi_\alpha(\rho_<)f_C^{(+)}(\rho_>)}
 {W[\varphi_\alpha,f_C^{(+)}]},
\end{equation}
and a localized source $\mathcal J$ generates the unit-outgoing-wave
coefficient
\begin{equation}
 \mathcal A_{\rm out}
 =-\frac{C}{W[\varphi_\alpha,f_C^{(+)}]}
 \int_0^\infty\varphi_\alpha(\rho)\mathcal J(\rho)\,\dd\rho.
 \label{eq:jost-amplitude}
\end{equation}
Since $\varphi_\alpha=\alpha u_k^{\rm reg}$, with $k^2=\omega^2-\Lam_{\rm f}$, and
$f_C^{(+)}=Cf_1^{(+)}$, the factors $\alpha$ and $C$ cancel against the
Wronskian.  Hence the leading radiation zero is determined solely by
\begin{equation}
 \mathcal F_2^{(2)}
 =\int_0^\infty u_{k_2}^{\rm reg}(\rho)\mathcal J_2(\rho)\,\dd\rho=0
 \label{eq:jost-overlap-zero}
\end{equation}
and cannot be generated by the right boundary or a Jost normalization.

Finite-domain calculations were repeated for a sequence of outer radii
$\rho_{\max}$ and radial spacings, with an $h^2\to0$ extrapolation.  We use
the superscript ${\rm FD}$ for the extrapolated finite-domain root
$\dlt^{\rm FD}$ and denote the independently obtained
half-line Jost--Green root by $\dlt^{\rm Jost}$.  The results for $\dlt_\star$ and $\dlt_\dagger$
are summarized in Table~\ref{tab:jost-two-zeros}.
For $\dlt_\star$, the Jost and finite-domain roots differ by
$1.10\times10^{-11}$; for $\dlt_\dagger$ they differ by only
$2.0\times10^{-12}$.  In both cases the overlap derivative is nonzero, so the
roots are simple rather than tangential zeros.  For the boundary test, define
$\Delta\dlt_{\rm root}(\rho_{\max})\equiv\dlt_{\rm root}(\rho_{\max})-\dlt_{\rm root}(\infty)$.
The maximum drift $|\Delta\dlt_{\rm root}|$ for $\rho_{\max}\ge25$ is below
$2\times10^{-11}$ for either root.

\begin{table*}[t]
 \centering
 \caption{Half-line Jost--Green validation of the two representative quadratic FGR zeros.
 The last column gives the maximum outer-radius drift for $\rho_{\max}\ge25$.
 Additional digits are retained only to document numerical agreement.}
 \label{tab:jost-two-zeros}
 \begin{tabular}{ccccc}
  \toprule
  root & $\dlt^{\rm Jost}$ & $\dlt^{\rm FD}$ &
  $\partial_\dlt\mathcal F_2^{(2)}$ & $\max|\Delta\dlt_{\rm root}|$ \\
  \midrule
  $\dlt_\star$ & 0.1241657536900 & 0.1241657536789 & $-1.6926$ & $1.6\times10^{-11}$ \\
  $\dlt_\dagger$ & 0.2271076778037 & 0.2271076778017 & $+1.0450$ & $1.9\times10^{-11}$ \\
  \bottomrule
 \end{tabular}
\end{table*}

Varying the regular and Jost normalizations and the matching radius leaves
the physical outgoing amplitudes unchanged at the quoted numerical precision.
Neither cancellation is generated by the finite outer boundary or by the
scattering normalization.

For the three additional thinner-wall roots, we independently solve the
complex sourced boundary-value problem by adaptive collocation.  We impose
$w_2(0)=0$ at the origin and the outgoing condition
$w_2'(\rho_{\max})=\ii k_2w_2(\rho_{\max})$ at $\rho_{\max}=40$.  An
inward-integrated outgoing Jost solution supplies the Wronskian phase.  The root of
$\operatorname{Re}[-W\mathcal A_{\rm out}]$ agrees with the signed-overlap
root on the same $h=0.005$ grid to within $1.4\times10^{-12}$
(Table~\ref{tab:jost-additional-zeros}).  The outgoing amplitudes obtained
by collocation and by Eq.~\eqref{eq:jost-amplitude} differ by less than
$3.3\times10^{-12}$ at the roots and at detunings $\pm10^{-5}$, where
the signed amplitudes reverse sign.  These checks use a shared bubble and
bound eigenfunction, but independent outgoing-response solvers; they are
not a separate continuum extrapolation of the three roots.

\begin{table*}[t]
 \centering
 \caption{Additional outgoing-BVP/Jost checks at $h=0.005$ and
 $\rho_{\max}=40$. The third column compares with the signed-overlap
 root on the same grid. The last column gives the maximum relative
 Wronskian variation at matching radii $2,5,10,20,30,40$, over the root
 and neighboring calculations. Extra digits document agreement only.}
 \label{tab:jost-additional-zeros}
 \begin{tabular}{cccc}
  \toprule
  root & $\dlt^{\rm out}$ & $|\dlt^{\rm out}-\dlt^{\rm overlap}|$
       & $\max|\Delta W|/|W|$ \\
  \midrule
  $\dlt_a$ & 0.0551760034146 & $7.7\times10^{-14}$ & $6.3\times10^{-11}$ \\
  $\dlt_b$ & 0.0675986603360 & $1.2\times10^{-13}$ & $6.1\times10^{-11}$ \\
  $\dlt_c$ & 0.0873520278803 & $1.4\times10^{-12}$ & $5.1\times10^{-11}$ \\
  \bottomrule
 \end{tabular}
\end{table*}

\paragraph{Threshold-preserving robustness of the two representative zeros.}
To isolate wall-interior changes from threshold motion, we introduce the
dimensionless deformation parameter $\epsilon$ and deform only the wall
interior using
\begin{equation}
 \widehat V_\epsilon^{(c)}(s;\dlt)
 =\widehat V(s;\dlt)+\epsilon s^3(1-s)^3(2s-1).
 \label{eq:curvature-preserving-deformation}
\end{equation}
If $f_c(s)=s^3(1-s)^3(2s-1)$, then
$f_c=f_c'=f_c''=0$ at both $s=0$ and $s=1$.  The two vacuum positions, their
free-energy difference, and both vacuum curvatures are therefore unchanged at
fixed $\dlt$.  In particular, the false-vacuum continuum threshold remains
$\Lam_{\rm f}=4-12\dlt$ and cannot itself generate a moving zero.

A fixed-grid scan at Hessian spacing $h=0.005$ tracks both roots for
$-0.02\le\epsilon\le0.02$ in steps of $0.01$.  Table~\ref{tab:curvature-preserving}
shows that both roots persist, remain simple, and stay separated by about
$0.103$ throughout the scan.  We denote this separation by
$\Delta\dlt_{\dagger\star}\equiv\dlt_\dagger-\dlt_\star$.  The overlap slopes remain well away from zero,
and the open-channel wave numbers remain of order unity.  Thus the two zeros
survive even when the vacuum curvatures and continuum thresholds are held
fixed; their motion is genuinely associated with changes in the wall and
scattering structure.

\begin{table*}[t]
 \centering
 \caption{Vacuum-curvature-preserving robustness scan using
 Eq.~\eqref{eq:curvature-preserving-deformation} at $h=0.005$.  The final
 column gives the separation of the two representative simple FGR zeros.}
 \label{tab:curvature-preserving}
 \begin{tabular}{cccccc}
  \toprule
  $\epsilon$ & $\dlt_\star$ &
  $\partial_\dlt\mathcal F_{2,\star}^{(2)}$ &
  $\dlt_\dagger$ &
  $\partial_\dlt\mathcal F_{2,\dagger}^{(2)}$ &
  $\Delta\dlt_{\dagger\star}$ \\
  \midrule
  $-0.02$ & 0.124216 & $-1.6632$ & 0.227794 & $+1.0145$ & 0.10358 \\
  $-0.01$ & 0.124191 & $-1.6779$ & 0.227450 & $+1.0297$ & 0.10326 \\
  $0$     & 0.124166 & $-1.6926$ & 0.227108 & $+1.0450$ & 0.10294 \\
  $+0.01$ & 0.124139 & $-1.7073$ & 0.226768 & $+1.0604$ & 0.10263 \\
  $+0.02$ & 0.124111 & $-1.7222$ & 0.226430 & $+1.0760$ & 0.10232 \\
  \bottomrule
 \end{tabular}
\end{table*}

\paragraph{Small-amplitude real-time validation at $\dlt_\dagger$.}
The root at $\dlt_\dagger$ is also visible directly in real-time constrained dynamics.
We run the projected radial PDE at $A=0.08$ and $0.10$ at
$\dlt_\dagger=0.2271076778$ and at symmetric offsets
$\Delta\dlt=\pm10^{-3}$.  The effective coefficient
$\Gamma_{2,\rm eff}^{\rm PDE}=P_2/A^4$ develops a pronounced minimum at the
Jost-validated root; Table~\ref{tab:reference-dagger-pde-spot} summarizes the targeted validation.  At $A=0.08$ the second-harmonic coefficient is reduced by roughly
$7\times10^3$ relative to either detuned point, and at $A=0.10$ by roughly
$8\times10^2$--$9\times10^2$.  Meanwhile $P_3$ remains finite.  At the root
$P_2/P_3=8.94\times10^{-4}$ for $A=0.08$ and
$4.74\times10^{-3}$ for $A=0.10$.  The finite third-harmonic signal also
shows that the suppression is channel selective rather than a general loss
of radiation.

\begin{table*}[t]
 \centering
 \caption{Small-amplitude constrained-PDE validation near $\dlt_\dagger$.  The three $\Gamma_{2,\rm eff}^{\rm PDE}$ columns correspond
 to $\Delta\dlt=-10^{-3},0,+10^{-3}$, respectively.}
 \label{tab:reference-dagger-pde-spot}
 \begin{tabular}{ccccc}
  \toprule
  $A$ & $\Gamma_{2,-}^{\rm PDE}$ & $\Gamma_{2,0}^{\rm PDE}$ &
  $\Gamma_{2,+}^{\rm PDE}$ & $(P_2/P_3)_0$ \\
  \midrule
  0.08 & $2.4447\times10^{-6}$ & $3.5764\times10^{-10}$ &
  $2.5217\times10^{-6}$ & $8.94\times10^{-4}$ \\
  0.10 & $2.3186\times10^{-6}$ & $2.9315\times10^{-9}$ &
  $2.6166\times10^{-6}$ & $4.74\times10^{-3}$ \\
  \bottomrule
 \end{tabular}
\end{table*}

The unprojected and consistently projected third-harmonic calculations at
$\dlt_\star$ give, after the same second-order extrapolation,
\begin{equation}
 \Gamma_3^{\rm un}=2.60157\times10^{-7},
 \qquad
 \Gamma_3^\perp=2.33409\times10^{-7}.
 \label{eq:gamma3-value}
\end{equation}
At the finest grid,
$\langle u_-,\mathcal J_2\rangle=-7.4672\times10^{-3}$ and
$c_-=7.9542\times10^{-4}$.  Removing the corresponding localized component
of $w_2$ lowers the third-harmonic power coefficient by $10.3\%$.  The
projected coefficient is small, so the associated decay is much slower than
ordinary second-harmonic damping.  Table~\ref{tab:gamma3-projection}
summarizes the convergence.

\begin{table}[t]
 \centering
 \caption{Effect of the fixed negative-mode projection on the
 third-harmonic coefficient.  The last two columns are in units of
 $10^{-7}$.}
 \label{tab:gamma3-projection}
 \begin{tabular}{ccc}
  \toprule
  Hessian spacing $h$ & $\Gamma_3^{\rm un}$ & $\Gamma_3^\perp$ \\
  \midrule
  0.0100 & 2.60338 & 2.33578 \\
  0.0050 & 2.60202 & 2.33452 \\
  0.0025 & 2.60168 & 2.33420 \\
  $h\to0$ & 2.60157 & 2.33409 \\
  \bottomrule
 \end{tabular}
\end{table}

\subsection{Higher-harmonic radiation and center-stable validation}

A generic bubble-mode excitation radiates first through the second harmonic.
At the selected root $\dlt_\star$, that channel is parametrically demoted and
the third harmonic becomes asymptotically leading.  The radiation zero
therefore changes the decay hierarchy, not only the scattering amplitude.

At the ordinary point $\dlt=0.1$, a scan over $A=0.08$--$0.24$ gives
\begin{equation}
 P_2\propto A^{3.972},
 \label{eq:pde-a4-fit}
\end{equation}
in agreement with the expected fourth power.  At the smallest amplitude,
\begin{equation}
 \Gamma_2^{\rm PDE}=1.6663\times10^{-3},
 \label{eq:pde-gamma2}
\end{equation}
which differs from Eq.~\eqref{eq:reference-benchmark} by $0.95\%$.

\begin{figure*}[t]
 \centering
 \includegraphics[width=\textwidth]{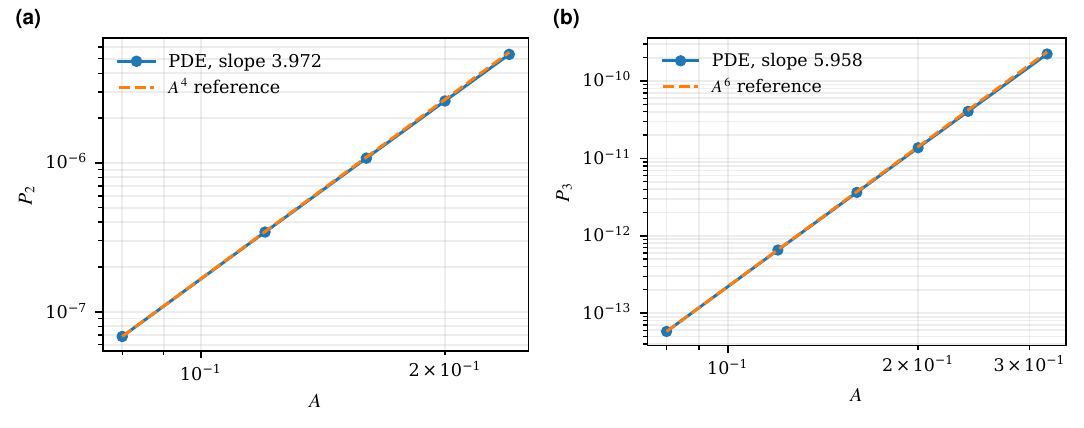}
 \caption{Direct real-time PDE tests of the change in the leading nonlinear
 radiation law. (a) At the ordinary point $\dlt=0.1$, the second-harmonic
 power has fitted exponent $3.972$, consistent with $P_2\propto A^4$.
 (b) At $\dlt_\star$, the third-harmonic power has fitted exponent $5.958$,
 consistent with $P_3\propto A^6$. Dashed lines show the corresponding
 expected scalings.}
 \label{fig:pde-power-scalings}
\end{figure*}

At $\dlt=\dlt_\star$, the third-harmonic power follows
\begin{equation}
 P_3\propto A^{5.958},
 \label{eq:pde-a6-fit}
\end{equation}
consistent with the sixth-power prediction.  The initial undressed scan
provides the corresponding power-law test.  The two amplitude laws are shown
together in Fig.~\ref{fig:pde-power-scalings}.  For a precision comparison
of the coefficient with the consistently projected outgoing-wave calculation,
we use the dressed and systematically converged evolution described below.

To suppress the start-up transient, we repeated the constrained evolution with
the consistently projected second-order initial data
\begin{equation}
 \begin{aligned}
 \chi(\rho,0)
 &=A u_{\rm sh}+A^2\left[w_0+\operatorname{Re}w_2^\perp\right],\\
 \partial_\tau\chi(\rho,0)
 &=2A^2\omega_{\rm sh}\,\operatorname{Im}w_2^\perp .
 \end{aligned}
 \label{eq:dressed-pde-initial-data}
\end{equation}
Here \(w_0\) and \(w_2^\perp\) are evaluated on the same discrete Hessian
grid as the real-time evolution.  The latter is obtained from the
sponge-matched linear problem
\begin{equation}
 \begin{aligned}
 \mathsf P_\perp
 \left[\mathcal H_0-4\omega_{\rm sh}^2
 -2\ii\omega_{\rm sh}\sigma(\rho)\right]w_2^\perp
 &=\mathsf P_\perp \mathcal J_2,\\
 \langle u_-,w_2^\perp\rangle&=0 .
 \end{aligned}
\end{equation}

We varied the spatial and temporal spacings, detector radius, sponge
location, strength and profile, and the extraction window.  Detector windows
were aligned at fixed retarded time using the third-harmonic group velocity.
The third-harmonic amplitude was obtained from a windowed simultaneous fit of
the fundamental through fourth harmonics; a rectangular window was rejected
because of visible leakage from the much larger bound-mode tail.

\begin{table}[t]
 \centering
 \caption{Convergence of the dressed projected-PDE third-harmonic
 coefficient at \(A=0.08\).  The time step is
 \(\Delta\tau=0.3\Delta\rho\).}
 \label{tab:dressed-gamma3-convergence}
 \begin{tabular}{ccc}
  \toprule
  \(\Delta\rho\) & \(\Delta\tau\) & \(\Gamma_3^{\rm PDE}\) \\
  \midrule
  0.040 & 0.0120 & \(2.25731\times10^{-7}\) \\
  0.030 & 0.0090 & \(2.29385\times10^{-7}\) \\
  0.020 & 0.0060 & \(2.31591\times10^{-7}\) \\
  0.015 & 0.0045 & \(2.32322\times10^{-7}\) \\
  0.010 & 0.0030 & \(2.33039\times10^{-7}\) \\
  \(\Delta\rho,\Delta\tau\to0\) &
  & \(2.33474\times10^{-7}\) \\
  \bottomrule
 \end{tabular}
\end{table}

A joint fit to the nine combinations
\(\Delta\rho=0.02,0.015,0.01\) and \(A=0.04,0.06,0.08\),
\begin{equation}
 \Gamma_3(\Delta\rho,A)
 =\Gamma_{3,0}
 +c_\rho(\Delta\rho)^2+c_AA^2
 +c_{\rho A}(\Delta\rho)^2A^2,
\end{equation}
where $c_\rho$, $c_A$, and $c_{\rho A}$ are fit coefficients.  The fit gives
\begin{equation}
 \Gamma_{3,0}^{\rm PDE}=2.33377\times10^{-7}.
\end{equation}
This differs by \(0.014\%\) from the independently extrapolated projected
outgoing-wave value
\begin{equation}
 \Gamma_3^\perp=2.33409\times10^{-7}.
\end{equation}
Allowing for the spread among the tapered extraction windows, we quote the
conservative result
\begin{equation}
 \Gamma_3^{\rm PDE}=2.334(9)\times10^{-7},
\end{equation}
which agrees with projected perturbation theory well within the numerical
uncertainty.  The retarded-time detector-radius spread is \(0.030\%\), the
sponge variation is \(0.010\%\), and the pre-reflection time-window variation
is \(0.056\%\).

As an independent check of the same second-order dressing, the projected
Poincar\'e--Lindstedt calculation gives
\(\omega_{\rm sh}^{(2)}=-9.7477\times10^{-3}\), while the measured \(A^2\) slope of
\(\omega_{\rm sh}(A)-\omega_{\rm sh}\) is \({-9.7504\times10^{-3}}\), a relative difference of
\(0.028\%\).

Across these scans, the largest residual negative-mode projection is below $2.0\times10^{-16}$.  The observed radiation therefore does not arise from numerical leakage into the unstable direction.

\label{subsec:center-stable-results}

The perturbative graph of the center-stable manifold provides an independent
check of the shooting construction.  We denote the tuned shooting value by
$\alpha_*(A)$.  Let $\mathcal J_1^{\rm un}$ and
$\mathcal J_3^{\rm un}$ denote the unprojected fundamental and third-harmonic
cubic sources and define
\begin{equation}
 c_{1,-}=\frac{\langle u_-,\mathcal J_1^{\rm un}\rangle}
 {\Lam_- -\Lam_{\rm sh}},
 \qquad
 c_{3,-}=\frac{\langle u_-,\mathcal J_3^{\rm un}\rangle}
 {\Lam_- -9\Lam_{\rm sh}}.
\end{equation}
Then
\begin{align}
 \alpha_*(A)&=c_\alpha^{\rm CS}A^3+\Order(A^5),\\
 c_\alpha^{\rm CS}&=\frac12\left[
 \operatorname{Re}(c_{1,-}+c_{3,-})
 +\frac{\omega_{\rm sh}}{\kappa}
 \operatorname{Im}(c_{1,-}+3c_{3,-})\right].
 \label{eq:cs-cubic}
\end{align}
The continuum extrapolations give
\begin{equation}
 \begin{aligned}
 c_\alpha^{\rm CS,PDE}&=-1.36724\times10^{-3},\\
 c_\alpha^{\rm CS,pert}&=-1.36679\times10^{-3},
 \end{aligned}
 \label{eq:cs-coefficient}
\end{equation}
a relative difference of $0.033\%$.  The nonlinear frequency-shift
coefficient agrees within $0.041\%$.  Perturbing the tuned initial value
produces exponential departure with rate $0.542435$, compared with
$\sqrt{-\Lam_-}=0.542284$, a difference of $0.028\%$.

Using $A=0.06,0.08,0.10$, four principal radial spacings down to
$\Delta\rho=0.015$ plus a $\Delta\rho=0.01$ check, detector radii 12 and 15,
and Hann and Blackman windows, a joint $(\Delta\rho)^2$--$A^2$
extrapolation gives
\begin{equation}
 \Gamma_3^{\rm CS}=2.604(10)\times10^{-7}.
 \label{eq:gamma3-center-stable}
\end{equation}
Relative to the unrounded center-stable fit value, the unprojected value
$\Gamma_3^{\rm un}=2.60157\times10^{-7}$ is $0.077\%$ lower, while the
fixed-projector value $\Gamma_3^\perp=2.33409\times10^{-7}$ is $10.35\%$ lower.
Thus the original unconstrained
critical-bubble equation selects the unprojected third-harmonic coefficient
on its center-stable manifold.  Table~\ref{tab:gamma3-dynamics} summarizes
the comparison.

\begin{table*}[t]
 \centering
 \caption{Third-harmonic coefficients for fixed-projector and center-stable
 dynamics, in units of $10^{-7}$. Percentages are relative to the unrounded
 center-stable fit value.}
 \label{tab:gamma3-dynamics}
 \begin{tabular}{lcc}
  \toprule
  Calculation & $\Gamma_3$ & Comparison with center-stable PDE \\
  \midrule
  Fixed-projector perturbation & 2.33409 & $10.35\%$ lower \\
  Center-stable PDE & $2.604(10)$ & reference \\
  Unprojected perturbation & 2.60157 & $0.077\%$ lower \\
  \bottomrule
 \end{tabular}
\end{table*}

\subsection{Finite-amplitude radiation minimum and local scaling}
\label{subsec:finite-minimum}

A spectral FGR zero is defined in the $A\to0$ limit, whereas a physical wall
excitation has finite amplitude.  At finite $A$, the zero becomes a displaced
radiation minimum.  We test whether its position and width follow the local
FGR normal form.

\begin{figure*}[t]
 \centering
 \includegraphics[width=0.70\textwidth]{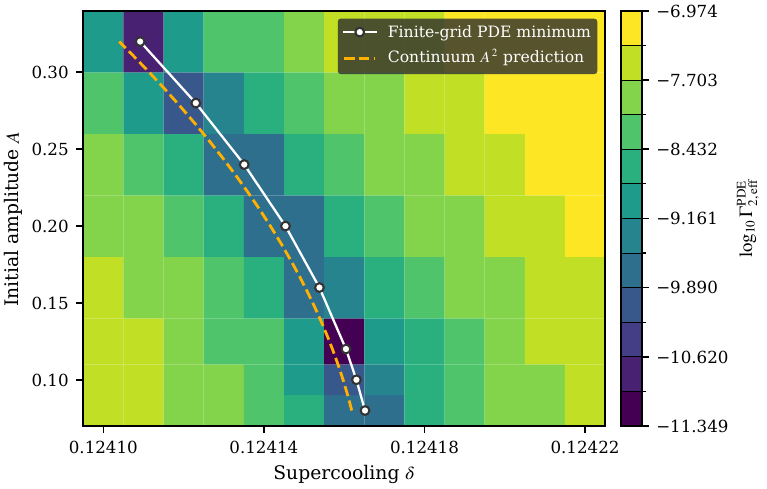}
 \caption{Zoomed view of the effective second-harmonic coefficient near the
 radiation valley in the supercooling--amplitude plane.  The dark minimum
 traces the finite-amplitude continuation of the $A\to0$ radiation zero; the
 overlaid points show the finite-grid PDE-extracted minimum at each amplitude.
 The dashed curve is the continuum fourth-order prediction in
 Eq.~\eqref{eq:perturbative-zero-curve}; finite-grid offsets are retained.}
 \label{fig:radiation-zero-heatmap}
\end{figure*}

A two-dimensional projected-PDE scan was first carried out over
\begin{equation}
 0.120\le\dlt\le0.128,
 \qquad
 0.08\le A\le0.32.
 \label{eq:2d-domain}
\end{equation}
The coarse scan contains 136 evolutions.  To resolve the radiation minimum,
we then performed 122 high-resolution runs in the narrow interval surrounding
the valley.  Figure~\ref{fig:radiation-zero-heatmap} shows this refined
region rather than the full coarse domain, so that the amplitude-dependent
shift of the minimum is visible directly.

The fourth-order calculation of Sec.~\ref{subsec:fourth-order-theory} gives,
in the continuum limit,
\begin{align}
 \partial_\dlt\mathcal F_2^{(2)}
 &=-1.6926,\\
 \mathcal F_2^{(4),\perp}(\dlt_\star)
 &=(-1.0217\times10^{-3})
 -(2.6138\times10^{-6})\ii .
 \label{eq:fourth-order-values}
\end{align}
Thus
\begin{equation}
 \dlt_{\rm min}^\perp(A)
 =\dlt_\star-6.0359(2)\times10^{-4}A^2+\Order(A^4).
 \label{eq:perturbative-zero-curve}
\end{equation}
The small imaginary part is only $0.256\%$ of the real part.  Consequently,
for $A\ne0$ the continuation is strictly a second-harmonic radiation
\emph{minimum}, not an exact zero of the complex amplitude.  At the minimum
the irreducible second-harmonic power is $\Order(A^8)$, so the
$\Order(A^6)$ third harmonic dominates asymptotically.

The continuum extrapolations of the quadratic shift are
\begin{equation}
 c_2^\perp=-6.0359\times10^{-4},
 \qquad
 c_2^{\rm CS}=-4.4078\times10^{-4}.
 \label{eq:shift-two-dynamics}
\end{equation}
A dressed projected-PDE extraction using three detector radii, late
source-time windows, and Hann and Blackman tapers gives
\begin{equation}
 c_2^{\rm PDE}=(-6.043^{+0.009}_{-0.005})\times10^{-4},
 \label{eq:zero-shift-pde}
\end{equation}
in agreement with Eq.~\eqref{eq:perturbative-zero-curve} at about $0.1\%$.
The center-stable displacement is distinct, just as the third-harmonic
coefficient is distinct between the two dynamics.

\label{subsec:local-scaling}

Writing
\begin{equation}
 \Delta\dlt=\dlt-\dlt_\star,
\end{equation}
a simple FGR root has
$\mathcal F_2^{(2)}=(\partial_\dlt\mathcal F_2^{(2)})\Delta\dlt+
\Order((\Delta\dlt)^2)$.  Because the normalization factor relating this overlap to the outgoing amplitude is smooth and nonzero at the root, one
therefore expects
$\Gamma_2\propto|\mathcal F_2^{(2)}|^2\propto(\Delta\dlt)^2$.
The high-resolution outgoing-wave scan confirms this expectation for
$|\Delta\dlt|\le10^{-4}$,
\begin{equation}
 \Gamma_2\propto|\Delta\dlt|^{p_\dlt},
 \qquad
 p_\dlt\simeq2.0000,
 \qquad 1-R^2\simeq2\times10^{-10}.
 \label{eq:local-gamma2-exponent}
\end{equation}
A direct quadratic fit is
\begin{equation}
 \Gamma_2=9.1204\,(\Delta\dlt)^2+\Order((\Delta\dlt)^3),
 \label{eq:local-gamma2-law}
\end{equation}
while the overlap itself is
\begin{equation}
 \mathcal F_2^{(2)}(\dlt)
 =-1.6926\,\Delta\dlt+\Order((\Delta\dlt)^2),
 \label{eq:local-fgr-linear}
\end{equation}
consistent with the independent Jost derivative obtained above.

The same local law organizes the finite-amplitude minimum and its width.
For projected dynamics, write $C=9.1204$ and let $D\ge0$ denote the
irreducible fourth-order quadrature contribution to the power.  Locally,
\begin{align}
 P_2 &\simeq C A^4\bigl(\Delta\dlt-c_2^\perp A^2\bigr)^2+D A^8,
 \qquad P_3\simeq\Gamma_3^\perp A^6,
 \label{eq:local-power-normal-form}\\
 \frac{\dd A}{\dd\tau}
 &\simeq-\frac{C}{\Lam_{\rm sh}}
       \bigl(\Delta\dlt-c_2^\perp A^2\bigr)^2A^3
       -\frac{\Gamma_3^\perp}{\Lam_{\rm sh}}A^5
       -\frac{D}{\Lam_{\rm sh}}A^7.
 \label{eq:local-envelope-normal-form}
\end{align}
These expressions retain the leading detuning, displacement, and quadrature
terms; smooth higher-order dependence on $\Delta\dlt$ and $A$ is omitted.
The center-stable coefficients replace the projected ones for the original
unconstrained dynamics.  At fixed $\dlt=\dlt_\star$, $P_2=\Order(A^8)$
and $P_3=\Order(A^6)$.  At a fixed nonzero detuning, however, the $A^4$
second-harmonic loss eventually recovers as $A\to0$.  Thus the
inverse-fourth-root asymptote applies at the zero, while nearby detuned
trajectories can exhibit an intermediate third-harmonic regime before
returning to inverse-square-root decay.

To leading order the boundaries $P_2=P_3$ lie a distance
$\sqrt{\Gamma_3^\perp/C}\,A$ from the displaced minimum.  Their full
separation is
\begin{equation}
 \begin{aligned}
 \Delta\dlt_{P_3>P_2}
 &=2\sqrt{\frac{\Gamma_3^\perp}{9.1204}}\,A+\Order(A^3)\\
 &=3.1995\times10^{-4}A+\Order(A^3).
 \end{aligned}
 \label{eq:third-band-local-law}
\end{equation}

\begin{figure*}[t]
 \centering
 \includegraphics[width=\textwidth]{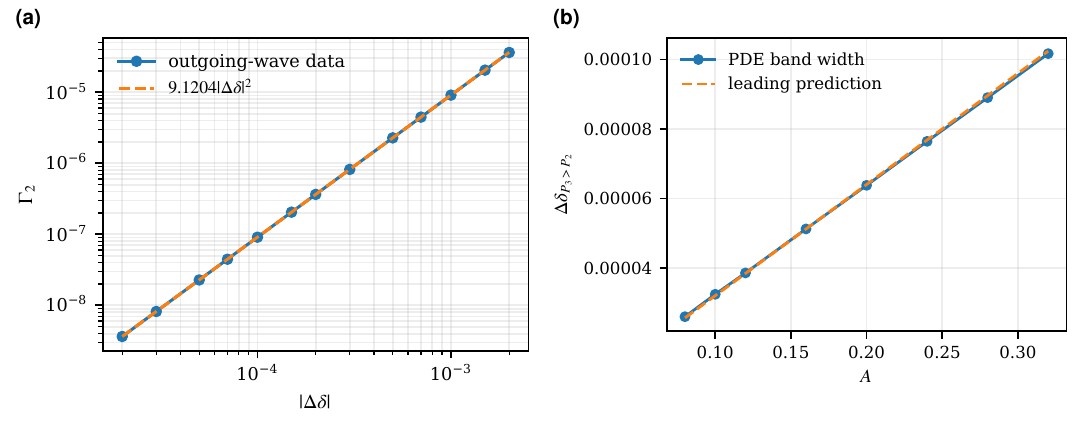}
 \caption{Local structure generated by the simple FGR zero, reconstructed
 from the high-resolution outgoing-wave and PDE data. (a) Quadratic
 suppression of the second-harmonic coefficient,
 $\Gamma_2\propto|\Delta\dlt|^2$. (b) Small-amplitude width of the region
 in which third-harmonic radiation dominates over the suppressed second
 harmonic.}
 \label{fig:local-scaling-summary}
\end{figure*}

The projected PDE gives a log--log exponent $0.9843$ with
$1-R^2\simeq1.6\times10^{-5}$.  Extrapolating $\Delta\dlt_{P_3>P_2}/A=K_0+K_2A^2$, with $K_0$ and $K_2$
as fit coefficients, to $A\to0$ gives
$K_0=3.2295\times10^{-4}$, only $0.94\%$ above the perturbative
coefficient.  Thus the simple-zero structure quantitatively organizes the
finite-amplitude two-dimensional radiation map, as summarized in
Fig.~\ref{fig:local-scaling-summary}.

\subsection{Power balance and slow-envelope reconstruction}

The interference zero changes the leading amplitude dependence of the energy
loss.  When the $A^4$ term disappears, the ordinary
$A^{-2}\propto\tau$ law crosses over to $A^{-4}\propto\tau$ as the third
harmonic becomes dominant.

At the ordinary point $\dlt=0.1$, using the unrounded source data for
Table~\ref{tab:radiation-scan}, Eq.~\eqref{eq:ordinary-decay} predicts
\begin{equation}
 \frac{\dd A^{-2}}{\dd\tau}
 =\frac{2\Gamma_2}{\Lam_{\rm sh}}
 =1.33974\times10^{-3}.
 \label{eq:ordinary-slope-prediction}
\end{equation}
A high-resolution run with $A_0=0.40$ gives
\begin{equation}
 \frac{\dd A^{-2}}{\dd\tau}
 =1.34689\times10^{-3},
 \label{eq:ordinary-slope-pde}
\end{equation}
a difference of $0.53\%$.  Fitting the envelope over the reported time
window to $C(\tau+\tau_0)^{-p}$ gives the effective exponent
\begin{equation}
 p=0.497.
 \label{eq:ordinary-exponent}
\end{equation}
The two initial amplitudes and the two grid resolutions give the same linear
$A^{-2}$ law within the finite-amplitude and discretization corrections
quantified below.  Table~\ref{tab:long-time-damping} summarizes the fitted
slopes, exponents, and goodness of fit.

At $\dlt=\dlt_\star$, the pure third-harmonic asymptotic prediction is
\begin{equation}
 \frac{\dd A^{-4}}{\dd\tau}
 =\frac{4\Gamma_3^\perp}{\Lam_{\rm sh}}
 =4.107\times10^{-7}.
 \label{eq:pure-third-slope}
\end{equation}
A high-resolution run with $A_0=0.32$ and final evolution time
$\tau_{\max}=6000$ gives a block-averaged linear fit
\begin{equation}
 \frac{\dd A^{-4}}{\dd\tau}
 =8.377\times10^{-7},
 \qquad
 1-R^2\simeq5.5\times10^{-5}.
 \label{eq:zero-direct-slope}
\end{equation}
At this finite amplitude and fixed $\dlt=\dlt_\star$, the
second-harmonic radiation-minimum curve has shifted, so a residual second
harmonic remains.  An independent short-time harmonic decomposition gives
\begin{equation}
 P_2=2.90\times10^{-10},
 \qquad
 P_3=2.33\times10^{-10},
 \label{eq:zero-powers}
\end{equation}
which predicts
\begin{equation}
 \left.\frac{\dd A^{-4}}{\dd\tau}\right|_{P_2+P_3}
 =8.562\times10^{-7}.
 \label{eq:zero-power-balance-slope}
\end{equation}
The carrier-resolved slope differs from this power-balance prediction by
$2.15\%$.  This agreement tests the local loss rate at finite amplitude;
the run does not span an order-one asymptotic decay.

\begin{table*}[t]
 \centering
 \caption{Carrier-resolved damping diagnostics. The reference for the ordinary run
 is the outgoing-wave prediction; for the finite-amplitude zero run it is the
 independently measured $P_2+P_3$ power-balance slope.}
 \label{tab:long-time-damping}
 \begin{tabular}{lcccc}
  \toprule
  case & transformed amplitude & PDE slope & reference slope & $1-R^2$ \\
  \midrule
  ordinary, $A_0=0.40$ & $A^{-2}$ & $1.34689\times10^{-3}$ & $1.33974\times10^{-3}$ & $3.3\times10^{-8}$ \\
  $\dlt_\star$, $A_0=0.32$ & $A^{-4}$ & $8.377\times10^{-7}$ & $8.562\times10^{-7}$ & $5.5\times10^{-5}$ \\
  \bottomrule
 \end{tabular}
\end{table*}

The strict $A\to0$ hierarchy is different: at fixed $\dlt_\star$ the residual second-harmonic power generated by finite-amplitude detuning is of higher order than $A^6$, whereas $P_3=\Gamma_3^\perp A^6$.  Therefore the third harmonic eventually dominates and Eq.~\eqref{eq:zero-decay} fixes the asymptotic law $A\sim\tau^{-1/4}$.  Its characteristic time,
\begin{equation}
 \tau_3=\frac{\Lam_{\rm sh}}{4\Gamma_3^\perp A_0^4},
 \label{eq:third-timescale}
\end{equation}
is $2.32\times10^8$ for $A_0=0.32$ and $1.52\times10^9$ for $A_0=0.20$.
The direct run to $\tau=6000$ is much shorter than these scales.  Its
measured slope corresponds to a fractional amplitude change of only about
$1.3\times10^{-5}$.  Consequently the inverse-fourth-root exponent is an
asymptotic prediction supported by measured radiation rates and the
reconstruction below, rather than a direct fit over an order-one
carrier-resolved amplitude decrease.

\begin{figure*}[t]
 \centering
 \includegraphics[width=\textwidth]{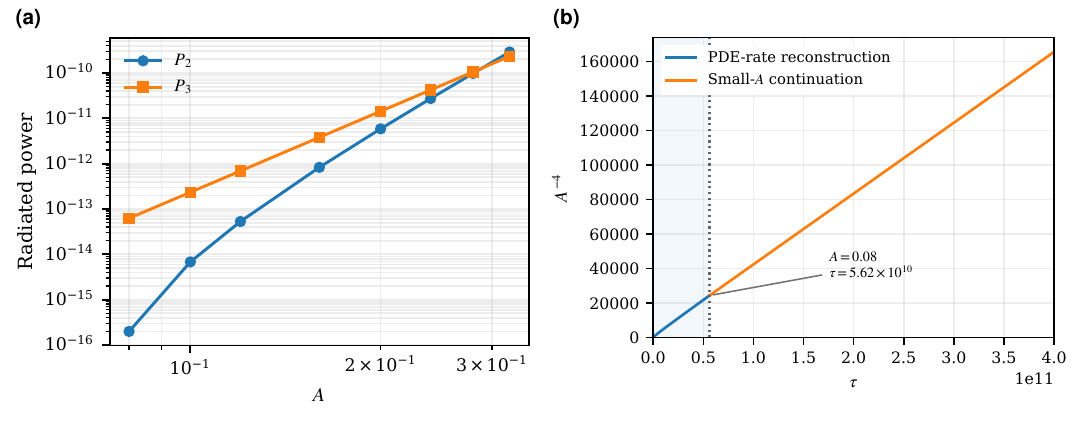}
 \caption{Slow-envelope reconstruction from PDE-measured powers at
 $\dlt_\star$. (a) Measured second- and third-harmonic powers along the
 amplitude sequence. (b) Integrated $A^{-4}$ slow envelope; the vertical
 line marks $A=0.08$ at $\tau=5.62\times10^{10}$, with blue denoting the
 PDE-measured-power reconstruction and orange the small-amplitude
 continuation. The reconstruction uses measured powers for
 $0.08\le A\le0.32$ and $P_3=\Gamma_3^{\rm PDE}A^6$ below $A=0.08$.
 These curves are not carrier-resolved PDE trajectories over the displayed
 time interval.}
 \label{fig:pde-slow-envelope}
\end{figure*}

To extend the fixed-projector test over an order-one amplitude change without
integrating every carrier oscillation over this long time scale, we use the
high-resolution two-dimensional projected-PDE scan to measure the harmonic
powers as functions of amplitude.  The complex second-harmonic amplitude and
third-harmonic power are interpolated at fixed
$\dlt=\dlt_\star$ for $0.08\le A\le0.32$ and inserted into the
cycle-averaged energy equation
\begin{equation}
 \Lam_{\rm sh} A\frac{\dd A}{\dd\tau}=-[P_2(A)+P_3(A)].
 \label{eq:pde-measured-envelope}
\end{equation}
At $A=0.32$ this procedure reproduces Eq.~\eqref{eq:zero-powers}.  The two
harmonics become equal at
\begin{equation}
 A_{P_2=P_3}=0.2895,
 \qquad
 \tau=5.81\times10^7,
 \label{eq:slow-envelope-crossover}
\end{equation}
and the third-harmonic fraction subsequently rises from $51.9\%$ at
$A=0.28$ to $82.1\%$ at $A=0.16$ and $99.68\%$ at $A=0.08$.  Starting from
$A_0=0.32$, the reconstructed envelope reaches
\begin{equation}
 \begin{aligned}
 A=0.16&\quad\text{at}\quad \tau=2.60\times10^9,\\
 A=0.08&\quad\text{at}\quad \tau=5.62\times10^{10}.
 \end{aligned}
 \label{eq:slow-envelope-times}
\end{equation}
Thus the interval supported directly by PDE-measured radiation rates covers a
factor-of-four reduction of the mode amplitude.  Below $A=0.08$ the measured
ratio is already $P_2/P_3=3.2\times10^{-3}$; continuing with the independently
converged small-amplitude PDE coefficient
$\Gamma_3^{\rm PDE}=2.334\times10^{-7}$ gives
\begin{equation}
 \frac{\dd A^{-4}}{\dd\tau}\longrightarrow
 \frac{4\Gamma_3^{\rm PDE}}{\Lam_{\rm sh}}=4.107\times10^{-7}.
 \label{eq:slow-envelope-asymptotic-slope}
\end{equation}
Figure~\ref{fig:pde-slow-envelope} shows the measured harmonic powers and
separates the reconstruction over the measured interval from the
small-amplitude continuation.

Appendix~\ref{app:slow-envelope} quantifies the much longer carrier-resolved
timescale of the asymptotic tail.

\section{Discussion}
\label{sec:discussion}

At $\dlt_\star$ and $\dlt_\dagger$, the second harmonic lies well inside the
false-vacuum continuum, with $k_2=2.5659$ and $k_2\simeq1.73$,
respectively, but its outgoing amplitude vanishes.  The cancellation occurs
between contributions from an extended source.  Its spatial organization is
different at the two representative points: a wall-scale inner--outer
partition at $\dlt_\star$, and alternating continuum-phase lobes at
$\dlt_\dagger$.  Both roots persist under a deformation that leaves the
vacuum positions, curvatures, and continuum thresholds unchanged.

The thinner-wall sign reversals show why this mechanism should not be
described by a global count of two roots.  In the thin-wall estimate the
bubble radius grows as $\rho\sim1/(3\dlt)$, so the phase sampled by the
extended source can change rapidly as $\dlt$ decreases.  This motivates the
inverse-supercooling scan but does not prove an infinite zero sequence.
Establishing the number or asymptotic spacing
of all zeros would require a controlled thin-wall scattering analysis.
Similarly, the positive samples beyond $\dlt_\dagger$ constrain the
resolved tail but do not establish a final zero on the localized branch.
The mode remains bound at the last resolved point, $\dlt=0.333$;
no spectral-merger location is inferred from the finite-box crossing.

The interference mechanism and the promotion of higher-harmonic radiation have
close precedents in oscillon decay
\cite{ZhangEtAl2020,CyncynatesGiurgicaTiron2021}.  Here we start from a
static Hessian eigenmode and follow its nonlinear continuation quantitatively.
Equation~\eqref{eq:local-power-normal-form} connects the
simple FGR zero to a minimum displaced by $c_2A^2$ and a
third-harmonic-dominated band of width proportional to $A$.  The finite
imaginary fourth-order amplitude leaves $P_2=\Order(A^8)$ even at the
minimum, whereas $P_3=\Order(A^6)$.  These relations predict both the
geometry of the radiation valley and the hierarchy of energy loss.  The
inverse-fourth-root law itself is the standard consequence of a leading
$A^6$ power; the model-specific information is how an open $A^4$ channel is
tuned away and how its finite-amplitude continuation is selected.

The unstable nature of the background is equally consequential.  Fixed
projection and center-stable tuning remove exponential departure in
different ways, and they need not preserve the same higher-order localized
response.  The measured center-stable value
$\Gamma_3^{\rm CS}=2.604(10)\times10^{-7}$ agrees with the unprojected
perturbative coefficient, while the projected value is about $10.35\%$
lower.  The displacement coefficients also differ,
$c_2^\perp=-6.0359\times10^{-4}$ and
$c_2^{\rm CS}=-4.4078\times10^{-4}$.
This comparison sharpens the relation to radiative damping on unstable
solitons \cite{LiLuhrmann2023,BizonRomanczukiewicz2026}: tuning the leading
FGR zero preserves the harmonic hierarchy, but the quantitative higher-order
dynamics depend on the treatment of the negative direction.

The slow-decay analysis separates two time scales.  Carrier-resolved PDE
evolution measures harmonic powers and tests finite-amplitude energy balance.
Integrating those measured rates reconstructs an order-one slow envelope,
with the third-harmonic fraction rising above $99\%$ at $A=0.08$.  This
reconstruction does not supply a carrier-resolved trajectory to
$\tau\sim10^{10}$--$10^{12}$.  Separating fast radiation measurements from
slow evolution also has precedents in oscillon longevity calculations
\cite{OllePujolasRompineve2021}; here the small-amplitude continuation is
fixed independently by the internal-mode radiation coefficient.

A kinematically open harmonic need not efficiently relax an excitation
maintained near a critical saddle.  This statement concerns intrinsic scalar
radiation, not the total lifetime of a generic expanding or collapsing
bubble.  Applications to nucleation dynamics must include the time dependence
of the wall, departure along the unstable direction, and any additional
dissipative channels.

\section{Conclusions}
\label{sec:conclusion}

The spherically symmetric internal mode of an $O(3)$ critical bubble admits
radiation zeros while its second harmonic remains kinematically open.  A
resolved signed-overlap scan identifies five roots on a finite parameter
interval, including three thinner-wall roots; it does not establish the
total count on the localized branch.  Detailed half-line, deformation, and
radial-PDE tests at two representative points establish distinct spatial
realizations of the same open-channel cancellation.

Near $\dlt_\star$, the radiation zero continues to a finite-amplitude
minimum with an $A^2$ displacement and an $A$-linear
third-harmonic-dominated width.  Radial evolution agrees with both
perturbative scalings.  The leading
third-harmonic power predicts $A\sim\tau^{-1/4}$ at the selected zero; measured
harmonic powers and slow-envelope reconstruction support the corresponding
relaxation hierarchy.  Center-stable shooting of the original unconstrained
equation selects the unprojected higher-order coefficient and distinguishes
physical saddle dynamics from fixed-projector evolution.

The radiation form factor of a static nucleation saddle controls both the
spectral cancellation and its nonlinear continuation.  Extension to evolving
bubble walls is required to assess its role in phase-transition dynamics.

\begin{acknowledgments}
The authors thank the participants of the Setouchi Summer Institute 2026 (SSI2026) for valuable discussions.
\end{acknowledgments}

\section*{Data Availability}

The numerical data and source code supporting the findings of this study, including the PDE evolution solver, are available from the corresponding author upon reasonable request.

\appendix

\section{Flux normalization}
\label{app:flux}

For a spherically symmetric perturbation, the dimensionless outward power through a sphere of radius $\rho$ is
\begin{equation}
 P(\rho,\tau)
 =-\rho^2\int \dd\Omega\,
 \partial_\tau\eta\,\partial_\rho\eta.
 \label{eq:flux-definition}
\end{equation}
Consider a single outgoing harmonic
\begin{equation}
 \eta_n
 =\frac{A^n}{\sqrt{4\pi}\rho}
 \operatorname{Re}\left[\mathcal A_ne^{\ii(k_n\rho-\omega_n\tau)}\right].
\end{equation}
At large radius, terms suppressed by $1/\rho$ relative to the phase derivative may be neglected.  Averaging over a cycle gives
\begin{equation}
 \Pbar_n
 =\frac{1}{2}\omega_n k_n|\mathcal A_n|^2A^{2n}.
 \label{eq:general-harmonic-power}
\end{equation}
Equations~\eqref{eq:gamma2} and \eqref{eq:gamma3} follow for $n=2$ and $n=3$.

\section{Numerical convergence and constraint diagnostics}
\label{app:convergence}

At the ordinary point $\dlt=0.1$, the long-time inverse-square slopes are
\begin{equation}
\begin{array}{c|c|c}
 A_0 & (\Delta\rho,\Delta\tau) & \dd A^{-2}/\dd\tau \\
 \hline
 0.40 & (0.04,0.012) & 1.34695\times10^{-3} \\
 0.25 & (0.04,0.012) & 1.34352\times10^{-3} \\
 0.40 & (0.02,0.006) & 1.34689\times10^{-3}
\end{array}
\end{equation}
compared with the outgoing-wave prediction $1.33974\times10^{-3}$.  The fine-grid relative difference is $0.53\%$.

For the two-dimensional scan, the maximum absolute negative-mode projection over all 258 evolutions is $1.97\times10^{-16}$.  For the long ordinary and radiation-zero runs it remains below $2.5\times10^{-16}$.  Halving the grid and time steps changes the leading ordinary coefficient by less than one percent.

The convergence diagnostics primarily document numerical precision and are
therefore tabulated.  The dressed
third-harmonic convergence is already given in
Table~\ref{tab:dressed-gamma3-convergence}.
Table~\ref{tab:zero-shift-convergence}
collects the grid convergence of the quadratic radiation-minimum displacement
for the fixed-projector and center-stable hierarchies.

\begin{table}[t]
 \centering
 \caption{Grid convergence of the quadratic radiation-minimum displacement
 coefficients. The final row is the $h\to0$ extrapolation.}
 \label{tab:zero-shift-convergence}
 \begin{tabular}{ccc}
  \toprule
  $h$ & $c_2^{\perp}$ & $c_2^{\rm CS}$ \\
  \midrule
  0.0200 & $-6.038279\times10^{-4}$ & $-4.409712\times10^{-4}$ \\
  0.0100 & $-6.036495\times10^{-4}$ & $-4.408245\times10^{-4}$ \\
  0.0050 & $-6.036049\times10^{-4}$ & $-4.407878\times10^{-4}$ \\
  0.0025 & $-6.035938\times10^{-4}$ & $-4.407787\times10^{-4}$ \\
  $h\to0$ & $-6.035900\times10^{-4}$ & $-4.407756\times10^{-4}$ \\
  \bottomrule
 \end{tabular}
\end{table}

The near-spinodal spectral check separates outer-domain truncation from
Hessian discretization. Table~\ref{tab:spectral-domain} shows that the
apparent crossing in a radius-25 box at $\dlt_{\rm box}\simeq0.33065$ disappears
as the box is enlarged. At this same parameter the converged binding ratio
is about $0.79897$, and at $\dlt=0.333$ it is about $0.79926$.
The last digits in the table record numerical convergence, rather than a
measurement of a continuum-merger parameter.

\begin{table*}[t]
 \centering
 \caption{Domain and grid checks of the near-spinodal shape-connected mode.
 The parameter $\dlt_{\rm box}\simeq0.33065$ identifies the old finite-box test point,
 not a physical merger. The continuum threshold is evaluated analytically.}
 \label{tab:spectral-domain}
 \begin{tabular}{cccc}
  \toprule
  $\dlt$ & $\rho_{\max}$ & $h$ & $\Lam_{\rm sh}/\Lam_{\rm f}$ \\
  \midrule
  $\dlt_{\rm box}$ & 25 & 0.010 & 0.99999793 \\
  $\dlt_{\rm box}$ & 50 & 0.010 & 0.80118464 \\
  $\dlt_{\rm box}$ & 100 & 0.010 & 0.79897215 \\
  $\dlt_{\rm box}$ & 200 & 0.010 & 0.79897146 \\
  $\dlt_{\rm box}$ & 200 & 0.020 & 0.79896786 \\
  $\dlt_{\rm box}$ & 200 & 0.005 & 0.79897236 \\
  0.333 & 100 & 0.010 & 0.82650211 \\
  0.333 & 200 & 0.010 & 0.79934107 \\
  0.333 & 400 & 0.010 & 0.79926122 \\
  0.333 & 600 & 0.010 & 0.79926122 \\
  0.333 & 600 & 0.020 & 0.79926077 \\
  0.333 & 600 & 0.005 & 0.79926134 \\
  \bottomrule
 \end{tabular}
\end{table*}

The terminal-time check uses the recovered unconstrained evolution code,
$\rho_{\max}=48$, a sponge beginning at $\rho=34$, and fixed
$\Delta\rho=0.02$, $\Delta\tau=0.006$. The requested horizons
$T=40,46,52$ are rounded to the nearest time step, giving
$40.002,46.002,52.002$. Table~\ref{tab:terminal-time} reports the tuned
shooting values and their variation relative to $T=46$. Because the
unstable tangent grows exponentially, the terminal residual alone is not
a useful measure of parameter accuracy; the supplied records retain both
the residual and its tangent sensitivity. This comparison concerns
$\alpha_*(A)$ at fixed discretization, not a new horizon extrapolation of
$\Gamma_3^{\rm CS}$.

\begin{table*}[t]
 \centering
 \caption{Terminal-time stability at fixed spatial and temporal steps.
 The last column is the largest relative change from $T=46$ over
 $T=40,46,52$, evaluated before rounding the displayed shooting values.}
 \label{tab:terminal-time}
 \begin{tabular}{ccc}
  \toprule
  $A$ & $\alpha_*(A;T=46)$ & $\max_T|\alpha_*(T)/\alpha_*(46)-1|$ \\
  \midrule
  0.06 & $-2.94952\times10^{-7}$ & $3.3\times10^{-10}$ \\
  0.08 & $-6.99338\times10^{-7}$ & $3.3\times10^{-10}$ \\
  0.10 & $-1.36629\times10^{-6}$ & $3.3\times10^{-10}$ \\
  \bottomrule
 \end{tabular}
\end{table*}

\section{Slow-envelope reconstruction and carrier-resolved scale}
\label{app:slow-envelope}

The slow-envelope reconstruction uses the high-resolution two-dimensional PDE
scan with $\Delta\rho=0.02$ and $\Delta\tau=0.006$.  For each sampled amplitude
$A=0.08$--$0.32$, the complex second-harmonic amplitude is interpolated to
$\dlt=\dlt_\star$ before converting it to $P_2$, while $P_3$ is
interpolated directly at the same fixed $\dlt$.  A shape-preserving cubic interpolant of the logarithm of each measured
power as a function of $\log A$ is then inserted into
Eq.~\eqref{eq:pde-measured-envelope}.  No perturbative coefficient is used
inside the directly measured interval.  Below $A=0.08$ the continuation uses
$P_3=\Gamma_3^{\rm PDE}A^6$ with
$\Gamma_3^{\rm PDE}=2.334\times10^{-7}$ and neglects $P_2$ because the
measured ratio is already $P_2/P_3=3.2\times10^{-3}$ at the matching point.

The multiscale reconstruction measures local radiation rates from the
projected radial PDE and integrates only the slow energy balance.  A direct
simulation would instead have to resolve every oscillation over the
asymptotic decay time.
Using the measured small-amplitude coefficient, the pure-tail change
$A=0.08\to0.04$ corresponds to $\Delta\tau\simeq8.9\times10^{11}$, or
approximately $1.5\times10^{14}$ explicit updates at the production time
step.  The slow-envelope result should
therefore be read as a slow-envelope reconstruction from PDE-measured
rates, with a small-amplitude continuation, rather than a carrier-resolved
trajectory over the same interval.

\section{Reproducible numerical protocol}
\label{app:protocol}

The bubble equation is solved as a singular boundary-value problem by adaptive collocation.  For the refined root-region scan a thin-wall seed centered at $\rho\simeq1/(3\dlt)$ is used, with outer radius
\begin{equation}
 \rho_{\max}=\max\left(25,\frac{1}{3\dlt}+20\right).
\end{equation}
The residual tolerance is $2\times10^{-9}$. For the corrected thick-wall samples
$0.25\le\dlt\le0.333$ and the representative values in
Table~\ref{tab:radiation-scan}, the domain is instead
$\rho_{\max}=\max(40,30/\sqrt{\Lam_{\rm f}})$ and the bubble residual
tolerance is $2\times10^{-10}$. A stable near-spinodal parametrization uses
$\epsilon=1-3\dlt$, $z=2\sqrt{\epsilon}\rho$, and
$s_b=\epsilon q/(2+\epsilon)$, so that
\[
 q''+\frac{2}{z}q'=q-q^2+\beta q^3,
 \qquad \beta=\frac{2\epsilon}{(2+\epsilon)^2}.
\]
Here primes denote $z$ derivatives, and
$\mathcal H_0/\Lam_{\rm f}=-\partial_z^2+1-2q+3\beta q^2$.
The Hessian is still evaluated on the declared physical radial grid.  The radial Hessian is discretized by a centered second-order difference and diagonalized as a symmetric tridiagonal matrix.  Bound modes are normalized with the discrete approximation to $\int u^2\dd\rho=1$.

For the outgoing problem, a regular homogeneous solution and a sourced particular solution are integrated simultaneously.  Their linear combination is fixed by $w_n'(\rho_{\max})=\ii k_nw_n(\rho_{\max})$.  Radiation zeros are located by root finding on the real regular-solution overlap in Eq.~\eqref{eq:fgr-overlap}.  The five roots in Table~\ref{tab:zero-convergence} were evaluated at
nominal $h=0.0100$, $0.0050$, and $0.0025$ and extrapolated to second order.
The uniform grid uses $N=\operatorname{round}(\rho_{\max}/h)$, so the
actual spacing is $\rho_{\max}/N$; the difference from nominal $h$ is
retained in the supplied numerical records.  The two reference roots
$\dlt_\star$ and $\dlt_\dagger$ were additionally checked with
half-line regular and Jost solutions. The three additional roots also
pass the fixed-grid outgoing-BVP/Jost comparison in
Table~\ref{tab:jost-additional-zeros}. The curvature-preserving scan in Eq.~\eqref{eq:curvature-preserving-deformation} tracks both roots at $h=0.005$ for the same five deformation values.  The cumulative profile at $\dlt_\dagger$ is evaluated on the $h=0.0025$ root at $\Delta\dlt_\dagger=0$ and $\pm2\times10^{-4}$, using the same regular continuum normalization as in Eq.~\eqref{eq:fgr-overlap}.

The refined scan uses 241 points uniform in $1/\dlt$ on
$0.05\le\dlt\le0.30$, supplemented by 86 points uniform in $\dlt$ on
$0.13\le\dlt\le0.30$ and seven local samples near each resolved root.
Removing duplicates gives 353 samples at nominal $h=0.005$.
The regular solution and its overlap integral are integrated with an
eighth-order explicit method, with relative and absolute tolerances
$10^{-10}$ and $10^{-12}$.  For each root, bracketing uses the signed
overlap, not a minimum of $\Gamma_2$.  The three additional thinner-wall
roots are also recomputed at $\rho_{\max}=40$.  The comparison uses the
same nominal spacing; because the actual grid spacing can change slightly,
the quoted difference combines domain and grid-rounding effects.
For Fig.~\ref{fig:gamma2-scan}, the samples at $\dlt\ge0.25$ are replaced
by 41 calculations on the enlarged domains defined above, giving 359
plotted samples on $0.05\le\dlt\le0.333$. The supplied assembly script
records the provenance of every row.

For real-time evolution, the production projected calculation uses a second-order velocity--Verlet update and enforces Eq.~\eqref{eq:constraint} after every velocity and field update.  The sponge profile is a fourth-power ramp.  The amplitude-scaling scans use delayed tapered demodulation after the outgoing pulse has reached the detector.  The validation at $\dlt_\dagger$ uses $\Delta\rho=0.04$, $\Delta\tau=0.012$, outer-domain radius $\rho_{\max}=70$, detector radius $\rho_{\rm det}=20$, and final time $\tau_{\max}=180$ at $A=0.08$ and $0.10$ for $\Delta\dlt_\dagger=0$ and $\pm10^{-3}$.  For the precision third-harmonic coefficient, the consistently projected data in Eq.~\eqref{eq:dressed-pde-initial-data} are constructed on the evolution grid, detector windows are aligned at fixed retarded time, and the fundamental through fourth harmonics are fitted simultaneously with Hann or Blackman windows.  The center-stable calculation evolves the original unconstrained PDE together with the tangent with respect to the shooting parameter and uses Newton continuation to impose Eq.~\eqref{eq:cs-terminal}.  The half-line boundary check independently constructs regular and outgoing Jost solutions and verifies Wronskian and normalization independence.  Carrier-resolved long-time envelopes are obtained by band-pass filtering around the internal frequency, constructing the analytic-signal envelope, and averaging over several oscillation periods.  The separate order-one slow-envelope reconstruction uses the PDE-measured harmonic powers as described in Appendix~\ref{app:slow-envelope}.

\bibliography{references}

\end{document}